\documentclass[a4paper,11pt]{article}
\usepackage{jcappub} % for details on the use of the package, please see the JINST-author-manual
\usepackage{lineno}
\usepackage{subcaption}
\usepackage{aas_macros}

\title{\boldmath Exploring Milky Way rotation curves with Gaia DR3: a comparison between $\Lambda$CDM, MOND, and General Relativistic approaches}

\author[a,b]{Beordo William,}
\affiliation[a]{INAF - Osservatorio Astrofisico di Torino, Via Osservatorio 20, 10025, Pino Torinese, Italy}
\affiliation[b]{Department of Physics, University of Turin, Via P. Giuria 1, 10125 Turin, Italy}
\emailAdd{william.beordo@inaf.it}
\author[a]{Crosta Mariateresa}
\emailAdd{mariateresa.crosta@inaf.it}
\author[a]{and Lattanzi Mario Gilberto}

\abstract{With the release of Gaia DR3, we extend the comparison between dynamical models for the Milky Way rotation curve initiated in the previous work. Utilising astrometric and spectro-photometric data for 719143 young disc stars within $|z|<1$ kpc and up to $R \simeq 19$ kpc, we investigate the accuracy of MOND and $\Lambda$CDM frameworks in addition to previously studied models, such as the classical one with a Navarro-Frenk-White dark matter halo and a general relativistic model. We find that all models, including MOND and $\Lambda$CDM, are statistically equivalent in representing the observed rotational velocities. However, $\Lambda$CDM, characterized by an Einasto density profile and cosmological constraints on its parameters, assigns more dark matter than the model featuring a Navarro-Frenk-White profile, with the virial mass estimated at $1.5\text{--}2.5 \times 10^{12} \, {\rm M}_{\odot}$ --- a value significantly higher than recent literature estimates. Beyond $10\text{--}15$ kpc, non-Newtonian/non-baryonic contributions to the rotation curve are found to become dominant for all models consistently. Our results suggest the need for further exploration into the role of General Relativity, dark matter, and alternative theories of gravitational dynamics in shaping Milky Way's rotation curve.}


\begin{document}
\maketitle
\flushbottom

\section{Introduction} \label{sec:intro}

With the advent of the new Gaia Data Release 3 (DR3), it is crucial to assess the accuracy of the dynamical models widely employed in the literature. Recently, carefully selected rotation curves of the Milky Way (MW) were presented by \cite{beordoGeometrydrivenDarkmattersustainedMilky2024e} for different stellar populations, using data from Gaia DR3. This work served as a follow-up of \cite{crostaTestingCDMGeometrydriven2020}, which utilised the previous Gaia DR2. Both studies focused on a direct comparison between a general relativistic velocity profile derived from the solution of \cite[BG, ][]{balasinNONNEWTONIANBEHAVIORWEAK2008} and a classical Newtonian model featuring a Navarro-Frenk-White dark matter halo (MWC). In the present work we extend the analysis of \cite{beordoGeometrydrivenDarkmattersustainedMilky2024e} to two state-of-the-art models in the context of galaxy dynamics: the MOdifyied Newtonian Dynamics \cite[MOND,][]{milgromModificationNewtonianDynamics1983} paradigm and the $\Lambda$CDM model, where the simulation-motivated Einasto profile \cite{Einasto1965, Navarro2004} is assumed to describe the CDM density distribution. This is the first time a fully general relativistic model is compared with the MOND approach and a classical model featuring an Einasto profile with $\Lambda$CDM constraints. The following sections provide a description of the data and the models adopted, as well as a discussion and comparison of the results with the MWC and BG models presented in \cite{beordoGeometrydrivenDarkmattersustainedMilky2024e}, along with other recent results in the literature.

\section{Data}
The data utilized in this study have been carefully selected from Gaia DR3, following the rigorous criteria outlined in the reference paper \cite{beordoGeometrydrivenDarkmattersustainedMilky2024e}. The resulting sample boasts high-quality astrometric and spectrophotometric information for a total of 719143 young disc stars located within $|z| < 1$ kpc and spanning from $R = 4.5$ to 19 kpc. This includes 241918 O-,B-,A-type stars (OBA), 475520 Red Giant Branch stars (RGB) with nearly-circular orbits, and 1705 classical Cepheids (DCEP), ensuring a comprehensive representation of various stellar tracers of the Galactic disc potential. 

Our sample does not extend beyond 19 kpc, as stringent quality criteria were applied in line with the Gaia Collaboration standards. Specifically, we required the relative uncertainty on the parallax to be smaller than 20\% for OBA and RGB stars. On the other hand, the sample of DCEP stars, provided by \cite{gaiacollaborationGaiaDataRelease2022}, comes with photometric distances with relative uncertainties smaller than 10\%. This selection ensures consistency in our analysis by focusing on highly reliable data, which is the priority for this study.

Leveraging this extensive sample, \cite{beordoGeometrydrivenDarkmattersustainedMilky2024e} derived rotation curves of the Milky Way for six distinct data sets: the 3 ‘pure’ data sets of OBA, DCEP, and RGB stars, the combined OBA + DCEP and RGB + DCEP samples, and the total sample consisting of all of the disc stars selected combined (OBA + RGB + DCEP, hereafter ALL). 

Here, in order to determine the error bars of the velocity profiles, the Robust Scatter Estimate (RSE) was adopted as a robust measure of the azimuthal velocity dispersion of the population in each radial bin, instead of performing the bootstrapping technique.\footnote{The RSE is defined as $(2\sqrt{2} {\,\rm erf}^{-1}(4/5))^{-1} \approx 0.390152 \times (Q90 - Q10) $, with Q90 and Q10 being the $90^{\rm th}$ and $10^{\rm th}$ percentiles of a distribution, and it coincides with the standard deviation in the case of a normal distribution.}
In fact, as discussed in \cite{beordoGeometrydrivenDarkmattersustainedMilky2024e}, the stellar azimuthal velocities are considerably dispersed around the median values (the RSE is typically $>10$ km s$^{-1}$), therefore smaller error bars would not encompass the actual variability of the sample. Moreover, with the large amount of stars per radial bin, a bootstrapped quantity would be much smaller than the individual uncertainty on the velocity measurements, representing thus a nonphysical situation.

In the present study, we employ these rotation curves to constrain the models outlined in sections~\ref{sec:dyn_mod}.

\subsection{Comments about the recent claim of a Keplerian rotation curve}
\label{sec:keplerian_fall}
Recent claims regarding a Keplerian rotation curve extending up to 30 kpc have been brought to light \cite{jiaoDetectionKeplerianDecline2023}, a finding seemingly consistent with various recent studies \cite{wangMappingMilkyWay2023, labiniMassModelsMilky2023, ouDarkMatterProfile2024, zhouCircularVelocityCurve2023, eilersCircularVelocityCurve2019}.
The radial distance range covered by the sample used in these studies extends almost 10 kpc beyond the range of our selected sample. These results, however, rely on methods and data selections that differ significantly from those adopted in this work.

The main distinctions emerge in the selection of the tracers. These studies typically use Luminous RGB stars \cite{zhouCircularVelocityCurve2023, ouDarkMatterProfile2024}, while in some cases the selected data sample is not specified \cite{wangMappingMilkyWay2023, labiniMassModelsMilky2023}. Our work, instead, uses a combination of OBA, RGB, and DCEP stars, and the derived rotation curves are consistent across these different tracers. In \cite{beordoGeometrydrivenDarkmattersustainedMilky2024e}, we imposed a stringent requirement of errors on parallaxes smaller than 20\% for OBA and RGB stars, in line with the Gaia Collaboration's accuracy standards, as larger uncertainties significantly reduce the reliability of stellar distance measurements. In contrast, the referenced literature employs various data--driven techniques that artificially assign smaller uncertainties than those measured, allowing them to extend the measured rotation curve to 30 kpc. For instance, \cite{wangMappingMilkyWay2023, labiniMassModelsMilky2023} adopt the Lucy’s Inversion Method, a Bayesian algorithm to reduce distance errors in the Gaia DR3 dataset beyond 20 kpc. Similarly, \cite{zhouCircularVelocityCurve2023, ouDarkMatterProfile2024} employ machine-learning models combining spectroscopy and photometry to infer “corrected” distances. Here, we abstain from delving into these techniques, acknowledging the reliability of the methods as asserted by the authors (see the discussion in the related works). Additionally, our work is the only one using DCEP stars, accurately selected by the Gaia Collaboration \cite{gaiacollaborationGaiaDataRelease2022}. This sample allows us to extend our rotation curve a bit farther out, as their distances are determined by photometric measurements (with relative uncertainty $< 10\%$). These are simply different methodological approaches. Our methodology strictly preserves the high quality of the data delivered by Gaia --- the most detailed individual object census ever --- without any modifications.

\begin{figure}[htbp]
    \centering
    \includegraphics[width=\linewidth]{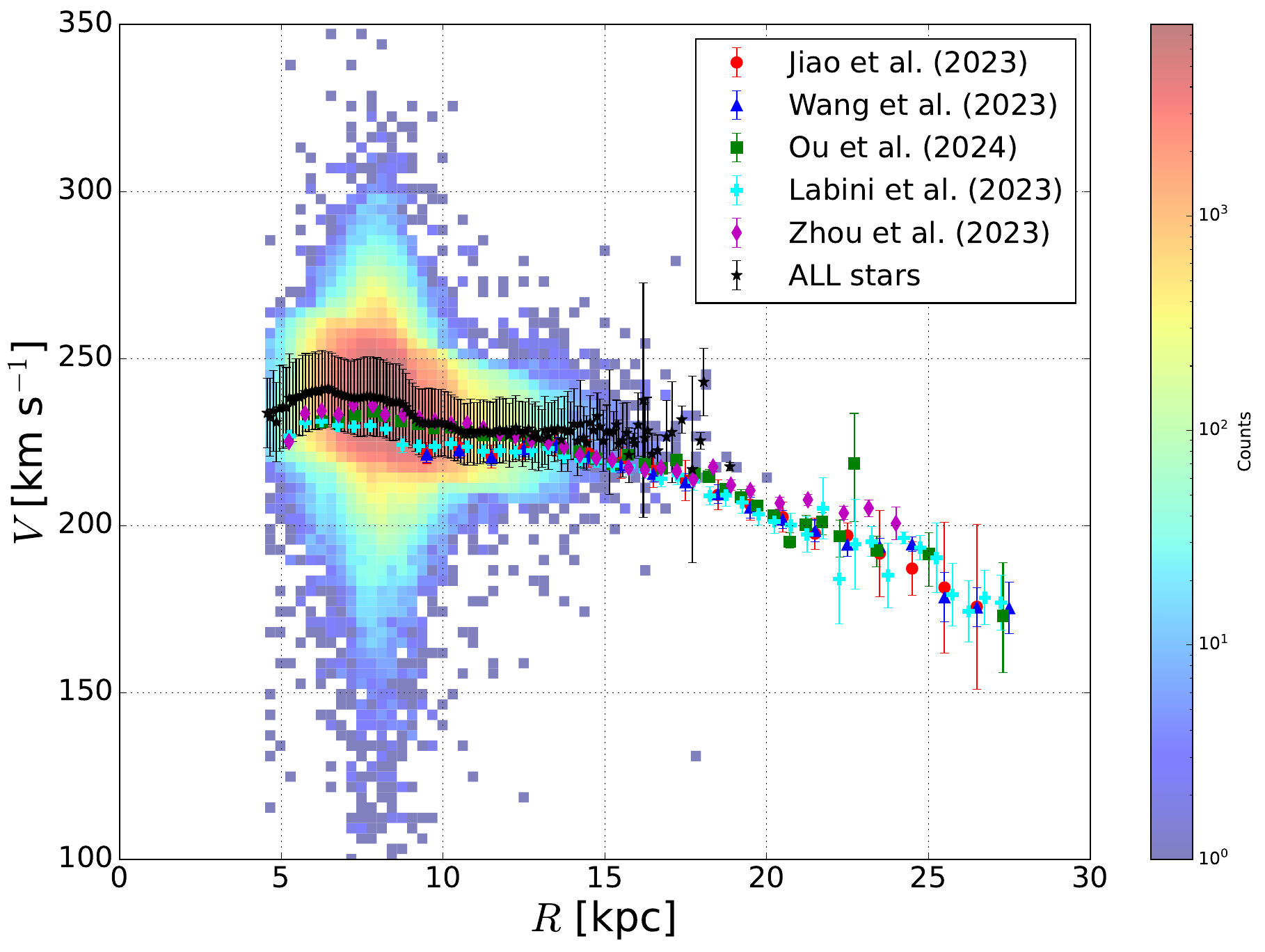}
    \caption{Rotation curve for the full sample, compared with rotation curves from the recent literature \cite{jiaoDetectionKeplerianDecline2023, wangMappingMilkyWay2023, ouDarkMatterProfile2024, labiniMassModelsMilky2023, zhouCircularVelocityCurve2023}. The 2D histogram of the full stellar sample in the $(R, V_\phi)$-space is plotted in the background \cite[see section 3 of][for more details about the data selection]{beordoGeometrydrivenDarkmattersustainedMilky2024e}.\label{fig:comparison}}
\end{figure}

Figure~\ref{fig:comparison} illustrates that our rotation curve a exhibit slightly declining profile, aligning with recent findings that indicate a pronounced decline only beyond 18--19 kpc;
the profiles of \cite{zhouCircularVelocityCurve2023} and \cite{ouDarkMatterProfile2024} are found to be the closest match. However, some differences also emerge within the overlapping range of 10--18 kpc. Specifically, our rotation curve shows a slight flattening around 15 kpc compared to other profiles (although statistical limitations of our selected sample start to become significant at that distance) and exhibits higher velocities in the inner region close to 5--7 kpc. These discrepancies may stem from other minor differences in the methodologies employed.

Firstly, while we refrained from conducting the Jeans analysis, being found negligible in the DR2 paper \cite{crostaTestingCDMGeometrydriven2020}, we instead implemented an eccentricity selection for the orbits of RGB stars. This adjustment was aimed at removing the effects of the asymmetric drift, to match the OBA and DCEP rotation curves. By conducting the Jeans analysis on our selected sample (as detailed in appendix~\ref{sec:jeans}), i.e., considering the derived circular velocity profile instead of the azimuthal one, the rotation curve shows a further slight increase within error bars (see figure~\ref{fig:jeans}), as expected. This suggests that the lack of the Jeans analysis in our procedure is unlikely to be the cause of the discrepancy observed at around 15 kpc.

Secondly, we delineated our study by confining our analysis to stars within $|z|<1$ kpc, deviating from the convention of considering a thicker disc encompassing $|z|<3$ kpc \cite{ouDarkMatterProfile2024}; for instance, a thinner disc selection reflects into higher velocities, especially between 5--15 kpc from the Galactic Centre \cite{jiaoDetectionKeplerianDecline2023, wangMappingMilkyWay2023}.

Thirdly, our depiction of error bars on the rotation curves incorporates the significant dispersion of azimuthal velocity among the stars. As shown in figure~\ref{fig:comparison}, the density plot highlights the azimuthal velocity distribution of our selected sample.
This stands in contrast to the rotation curves presented by the cited authors, whose error bars are derived by bootstrapping techniques and typically encompass only systematic sources of uncertainty, remaining relatively small compared to the velocity dispersion observed within each radial bin.

Lastly, we adopted quite small radial bins of $0.1$ kpc width (except for the DCEP sample for which we chose larger bins of $0.5$ kpc) compared to typical values of 0.5 and 1 kpc used in the literature. As discussed in appendix~\ref{sec:bins}, this of course plays an important role in the derived rotation curve, as the outer data points are distributed differently (see figure~\ref{fig:different_bins}). However, perfectly consistent results of the model parameters are obtained with either 0.5 or 1 kpc radial bins, indicating the robustness of the rotation curve defined in \cite{beordoGeometrydrivenDarkmattersustainedMilky2024e}.

\section{Dynamical models for the Milky Way}
\label{sec:dyn_mod}

\subsection{Baryonic component}
\label{sec:bar}
The distribution of baryonic matter in the Milky Way (MW) is modeled in the same way as in \cite{beordoGeometrydrivenDarkmattersustainedMilky2024e} for the MWC model. This includes a stellar bulge (b), a stellar thin (t) disc, and a stellar thick (T) disc (d). Both the MOND and $\Lambda$CDM models will use the same model for the baryonic component.

The bulge component is based on Plummer's density profile \citep{plummer}: \begin{equation} \rho_{\rm b} (r) = \frac{3 b_{\rm b}^2 M_{\rm b}}{4\pi(r^2 + b_{\rm b}^2)^{5/2}}, \label{eq
} \end{equation} where, in cylindrical coordinates, the bulge spherical radius is $r = \sqrt{R^2 + z^2}$, with $b_{\rm b}$ as the Plummer radius and $M_{\rm b}$ as the total bulge mass.

For the thin and thick MW discs, we adopt two stellar discs modeled using the Miyamoto-Nagai potential \cite{miyamotoNagai}: \begin{equation} \rho_{\rm d} (R,z)= \frac{M_{\rm d}b^2_{\rm d}}{4\pi} \frac{\Big[a_{\rm d}R^2 + \Big(a_{\rm d}+3\sqrt{z^2+b^2_{\rm d}}\Big)\Big(a_{\rm d}+\sqrt{z^2+b^2_{\rm d}}\Big)^2\Big]} {\Big[R^2\Big(a_{\rm d}+\sqrt{z^2+b^2_{\rm d}}\Big)^2\Big]^{5/2}\Big(z^2+b^2_{\rm d}\Big)^{3/2}}, \label{eq
} \end{equation} where $M_{\rm d}$ is the total mass of the disc (thin or thick), and $a_{\rm d}$ and $b_{\rm d}$ are the scale length and scale height, respectively.

Using the density profiles above, the Newtonian gravitational potential of the baryonic component is computed from Poisson's equation $\nabla^2 \Phi_{\rm bar} = 4\pi G(\rho_{\rm b} + \rho_{\rm td} + \rho_{\rm Td})$. The circular velocity is then obtained by solving the differential equation $V^2_{\rm bar}(R) = R \left(d\Phi_{\rm bar}/dR \right)$, which is equal to $V_{\rm b}^2 + V_{\rm td}^2 + V_{\rm Td}^2$, due to the linearity of Poisson's equation.

The free parameters of the baryonic matter distribution share the same prior distributions of the MWC model \cite[for details see sections 2 and B of][]{beordoGeometrydrivenDarkmattersustainedMilky2024e}.

\subsection{MOND}
\label{sec:mond}
In the MOND paradigm, the gravitational acceleration is
\begin{equation}
    \mathbf{g_{\rm MOND}} = \eta\left(\dfrac{g_{\rm N}}{g_0}\right) \mathbf{g_{\rm N}} \,,
	\label{eq:acc}
\end{equation}
where $\mathbf{g_{\rm N}}$ is the conventional Newtonian acceleration produced by the baryonic matter alone, while the interpolation function $\eta$ sets the transition between the Newtonian and the deep MOND regimes through the acceleration scale $g_0$. According to the MOND assumptions, when $g_{\rm N} \ll g_0$ the gravitational acceleration is boosted by a quantity $\eta \rightarrow \sqrt{g_0/g_{\rm N}}$ in order to obtain a flat rotation curve, while Newtonian dynamics is restored requiring $\eta \rightarrow 1$ when $g_{\rm N} \gg g_0$.
Here, we adopt the analytical expression proposed by \cite{mcgaughMilkyWayMass2008}, namely 
\begin{equation}
    \eta\left(\dfrac{g_{\rm N}}{g_0}\right) = \left(1 - e^{-\sqrt{g_{\rm N}/g_0}}\right)^{-1}\,.
	\label{eq:interp_nu}
\end{equation}
This interpolation function has been shown to provide an excellent representation of the Radial Acceleration Relation (RAR) observed for external disc galaxies \cite{lelliOneLawRule2017, mcgaughRadialAccelerationRelation2016}.

If we equal the gravitational acceleration $g_{\rm MOND}$ to the centripetal acceleration $g=V^2/R$ of an object orbiting in circular motion with velocity $V$ at a distance $R$ from the centre of a disc galaxy, the Mondian rotation velocity results
\begin{equation}
    V_{\rm MOND}(R, g_{\rm N}) = \sqrt{\dfrac{R g_{\rm N}}{1 - e^{-\sqrt{g_{\rm N}/g_0}}}}\,.
	\label{eq:v}
\end{equation}
Similarly, the magnitude of the Newtonian acceleration originated by the distribution of the baryonic matter alone can be written as $g_{\rm N} = V_{\rm bar}^2/R $; therefore the above expression becomes
\begin{equation}
    V_{\rm MOND}(R, V_{\rm bar}) = \dfrac{V_{\rm bar}}{\sqrt{1 - e^{-V_{\rm bar}/\sqrt{R g_0}}}}\,.
	\label{eq:v_mond}
\end{equation}

While in the MWC model of \cite{beordoGeometrydrivenDarkmattersustainedMilky2024e} the total rotation curve is given by adding in quadrature the dark matter halo component (namely $V_{\rm bar}^2 + V_{\rm h}^2$, where $V_{\rm bar}^2 = V_{\rm b}^2 + V_{\rm td}^2 + V_{\rm Td}^2$), in the MOND model the pure Mondian boost is represented by the denominator of equation~(\ref{eq:v_mond}) and writes explicitly as 
\begin{equation}
    V^{\rm MOND}_{\rm boost}(R, V_{\rm bar})
    =\sqrt{V_{\rm MOND}^2 - V_{\rm bar}^2}
    = V_{\rm bar} \sqrt{\eta(R, V_{\rm bar}) - 1}\, .
	\label{eq:v_mondBoost}
\end{equation}

As an additional parameter of the model, we have the acceleration scale $g_0$, which has been constrained to extremely tight values by the observed RAR of external galaxies \cite{lelliOneLawRule2017}, namely $g_0 = (1.20 \pm 0.02) \times 10^{-10}$ m s$^{-2}$. This scale is supposed to be fixed in the framework of MOND. However, given the small uncertainty on the parameter, setting a Gaussian prior $\mathcal{N}(\mu = 1.20, \sigma = 0.02) \times 10^{-10}$ m s$^{-2}$ does not affect the results \cite{liFittingRadialAcceleration2018}; therefore, strictly following the Bayesian approach, we prefer to not fix it and to marginalize over it afterwards.

\subsection{$\Lambda$CDM model with Einasto halo profile}
\label{sec:LCDM}
We consider the distribution of cold dark matter within the $\Lambda$CDM scenario to follow the Einasto density profile \cite{Einasto1965}, namely
\begin{equation}
    \rho_{\rm Einasto}(r) = \rho_{\rm s} \: \exp\left\{-\dfrac{2}{\alpha}\left[\left(\dfrac{r}{r_{\rm s}}\right)^{\alpha} - 1\right]\right\} \; ,
\end{equation}
where $\rho_{\rm s}$ is the characteristic density, $r_s$ the scale radius, and $\alpha$ the shape parameter.
Consistently with the prescriptions of \cite{liConstantCharacteristicVolume2019}, the parameters of the Einasto profile are written in terms of the halo concentration $C_{200} \equiv r_{200}/r_s$, where the virial radius $r_{200}$ is defined such that the enclosed average density is 200 times the critical density of the Universe, i.e. 
$\rho_{200} = 200 \rho_c = 75 H_0^2 / (\pi G)$, with $H_0 = 67$ km s$^{-1}$ Mpc$^{-1}$ \cite{planckcollaborationPlanck2013Results2014, duttonColdDarkMatter2014}; from here, the rotation velocity and the enclosed halo mass at the virial radius are then
\begin{equation}
    V_{200} = 10 C_{200} r_s H_0 \ ; \  M_{200} = \dfrac{V_{200}^3}{10 G^2 H_0^2} \ .
\end{equation}
With this redefinition, the following boundaries are set: $0<C_{200}<100$, $10<V_{200}$ [km s$^{-1}$] $<500$, and $0<\alpha<2$ \cite{liConstantCharacteristicVolume2019}. 
Three prior distributions, coming from $N$-body simulations within the $\Lambda$CDM cosmology, are then imposed to constrain the parameters:
\begin{itemize}
    \item Stellar Halo Mass (SHM) relation \cite{mosterGalacticStarFormation2013}:
            \begin{equation}
            \label{eq:SHM}
                \dfrac{M_\star}{M_{200}} = 2N \left[\left(\dfrac{M_{200}}{M_1}\right)^{-\beta} + \left(\dfrac{M_{200}}{M_1}\right)^{\gamma} \right]^{-1},
            \end{equation}
            where $M_\star = M_{\rm b} + M_{\rm td} + M_{\rm Td}$, $\log(M_1) = 11.59$, $N=0.0351$, $\beta=1.376$ and $\gamma=0.608$, with scatter $\sigma(\log M_{\star}) = 0.15$ dex.
    \item Halo mass-concentration relation \cite{maccioConcentrationSpinShape2008}:
            \begin{equation}
            \label{eq:HMC}
                \log(C_{200}) = a + b \log(M_{200}/[10^{12} h^{-1} {\rm M}_{\odot}]),
            \end{equation}
            where $a=0.977$ and $b=-0.130$ assuming the Planck cosmology, with a scatter of $0.11$ dex.
    \item Einasto shape parameter as a function of halo mass \cite{duttonColdDarkMatter2014}:
    \begin{equation}
    \label{eq:alpha}
        \alpha = 0.0095 \lambda^2 + 0.155,
    \end{equation}
     with a scatter of $0.16$ dex, where $\log(\lambda) = -0.11 + 0.146 m + 0.0138 m^2 + 0.00123 m^3$ and $m=\log(M_{200}/[10^{12} h^{-1} {\rm M}_{\odot}])$.
\end{itemize} 
Again, the baryonic matter distribution is modelled in the same way as the MOND and MWC models \cite[see section 2 of][]{beordoGeometrydrivenDarkmattersustainedMilky2024e}.

\section{Results}

Table~\ref{table:fit_params} lists, as best-fit estimates, the medians of the posteriors and their $1\sigma$ credible intervals. Details about the Bayesian analysis can be found in \cite{beordoGeometrydrivenDarkmattersustainedMilky2024e}.

\begin{figure}[htbp]
    \centering
    \includegraphics[width=\linewidth]{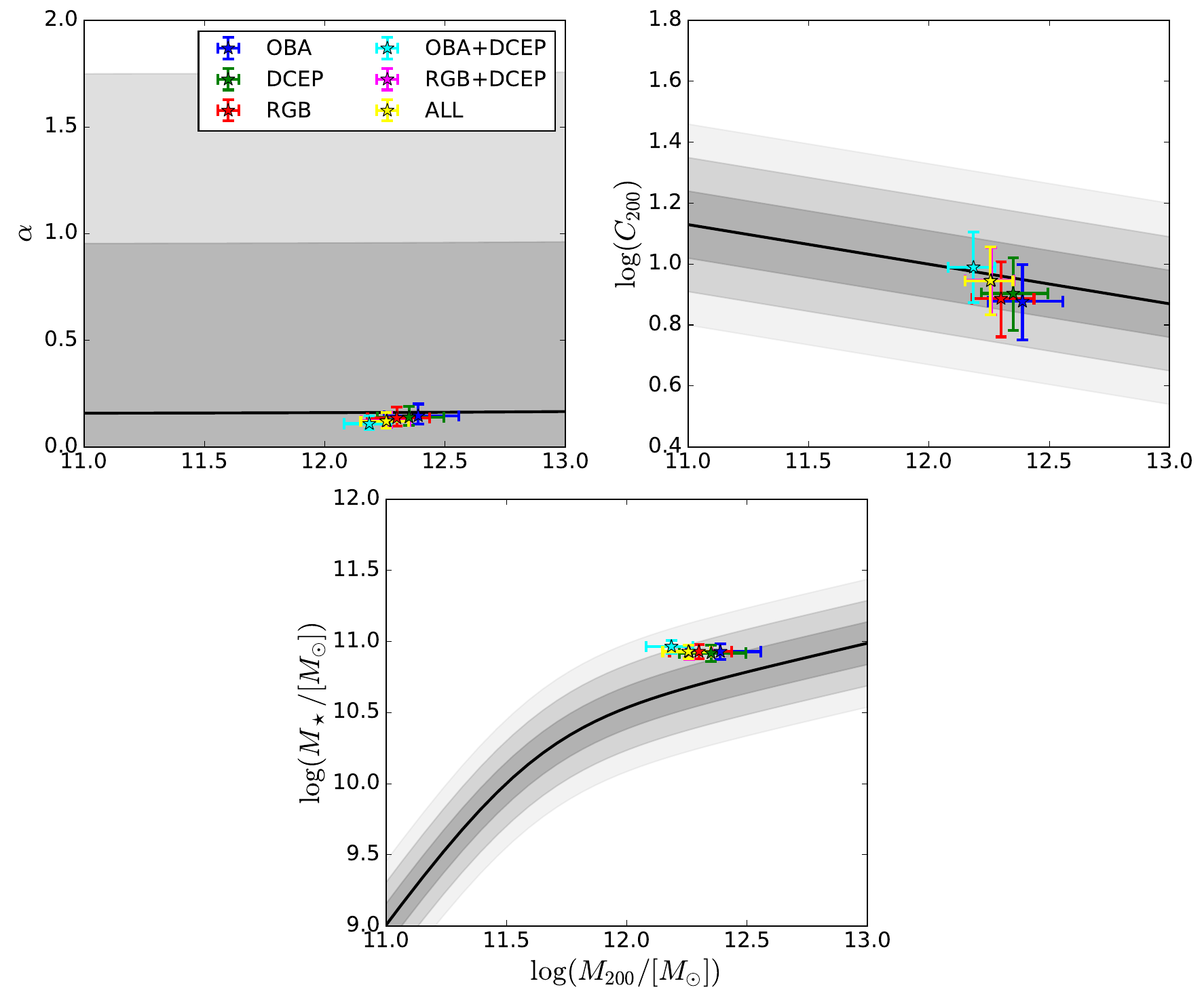}
    \caption{$\Lambda$CDM constraint relations, given by equations~(\ref{eq:SHM}),~(\ref{eq:HMC}), and~(\ref{eq:alpha}). Coloured data points are the best parameter estimates for each stellar sample (see table~\ref{table:fit_params}). The shaded gray regions represent the 1-,2-,3-$\sigma$ ranges around the mean relations.\label{fig:LCDMpriors}}
\end{figure}

\begin{figure}[htbp]
   %\centering % Not needed
   \begin{subfigure}{.495\linewidth}
        \includegraphics[width=\linewidth]{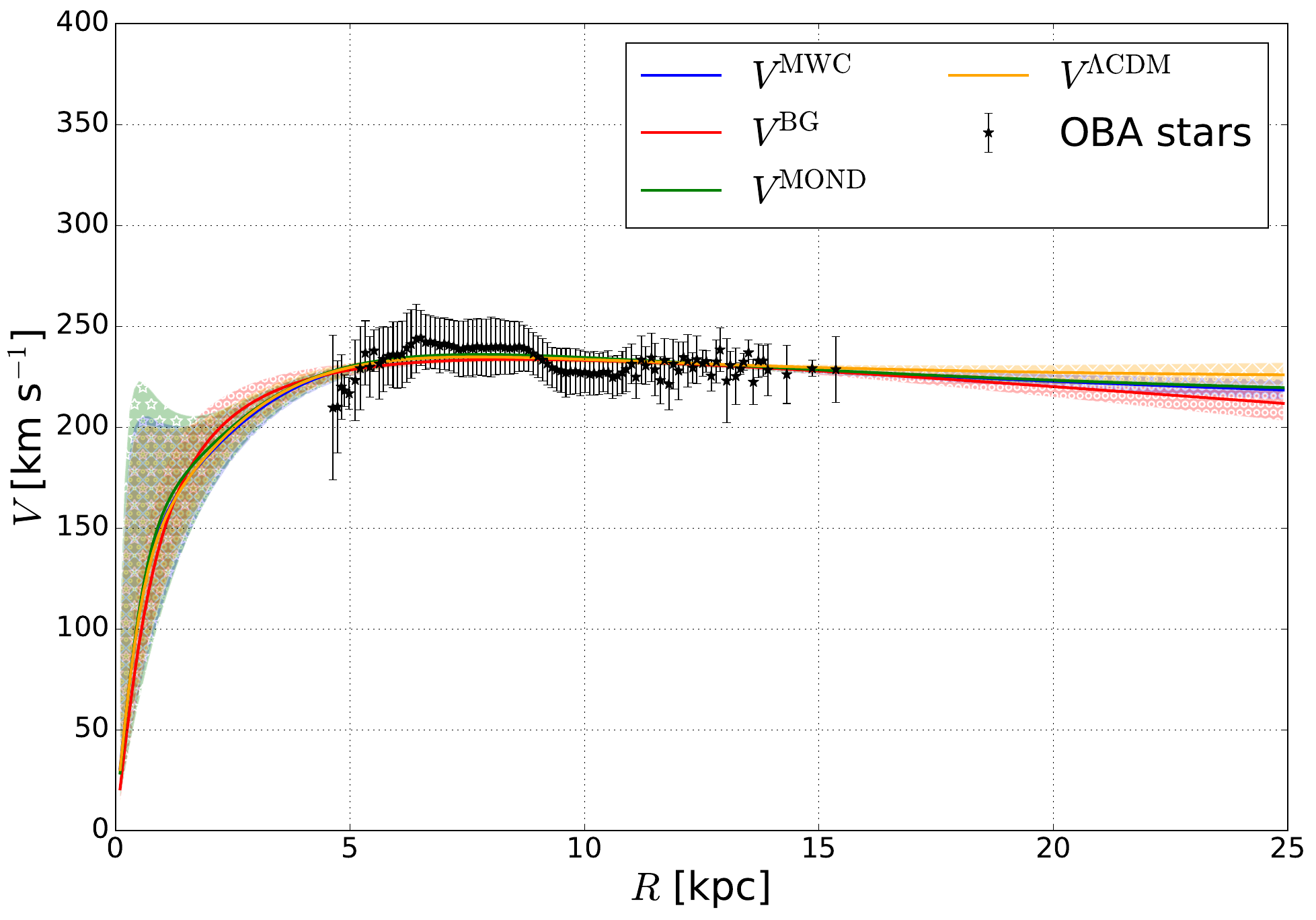}
    \end{subfigure}
    \hfill
    \begin{subfigure}{.495\linewidth}
        \includegraphics[width=\linewidth]{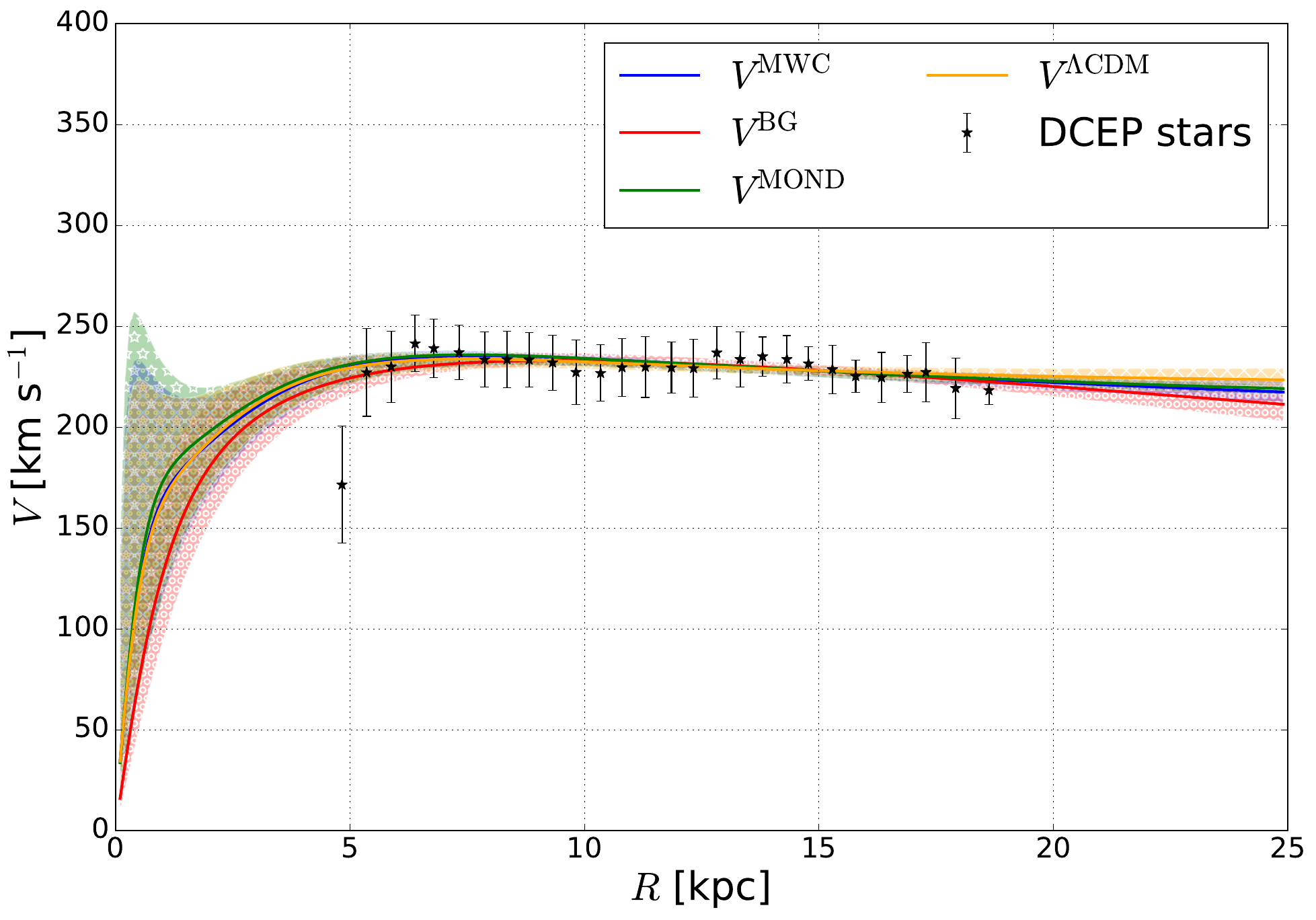}
    \end{subfigure}
    %% leave a blank line to create a line break

    \begin{subfigure}{.495\linewidth}
        \includegraphics[width=\linewidth]{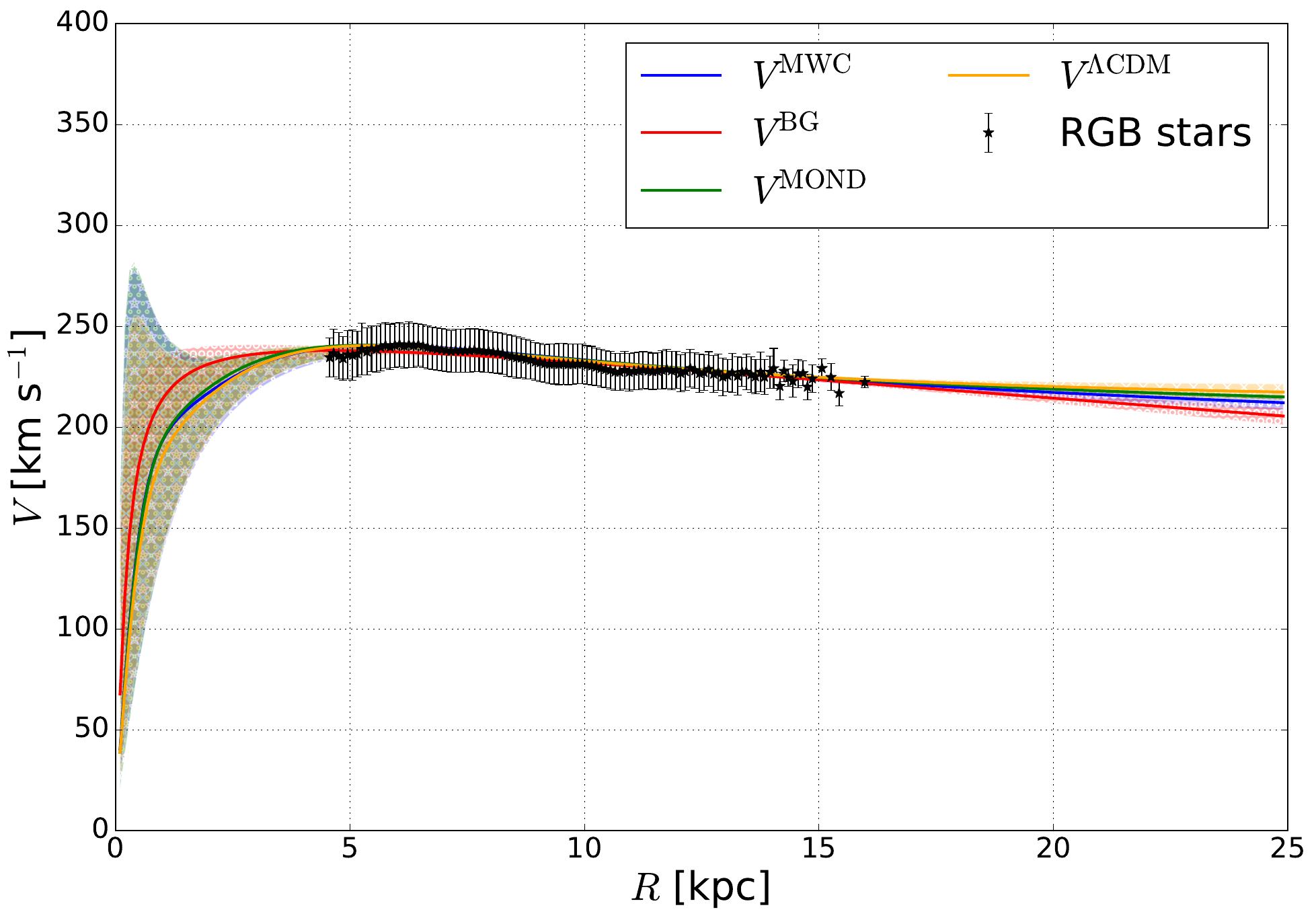}
    \end{subfigure}
    \hfill
    \begin{subfigure}{.495\linewidth}
        \includegraphics[width=\linewidth]{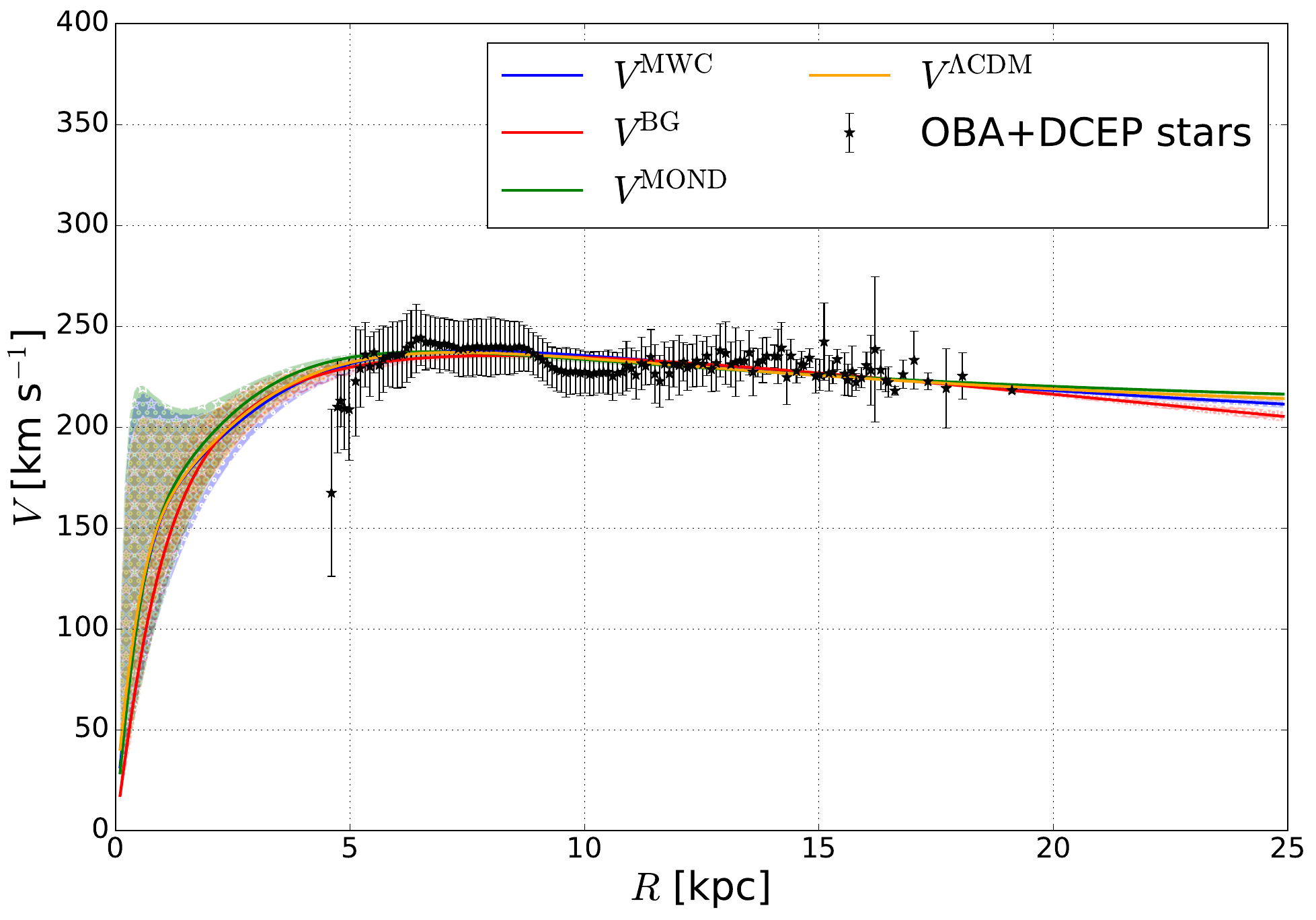}
    \end{subfigure}
     %% leave a blank line to create a line break

    \begin{subfigure}{.495\linewidth}
        \includegraphics[width=\linewidth]{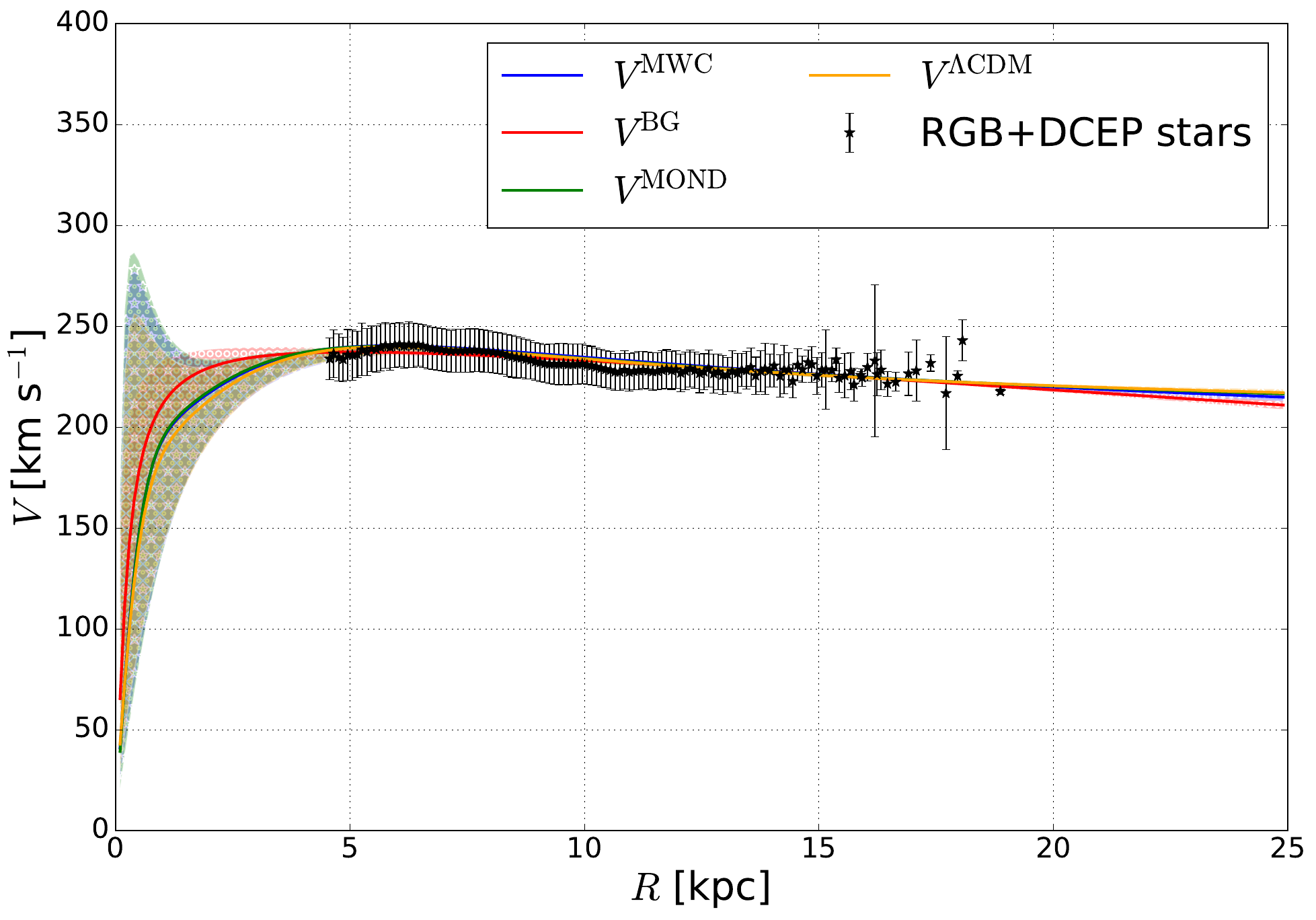}
    \end{subfigure}
    \hfill
    \begin{subfigure}{.495\linewidth}
        \includegraphics[width=\linewidth]{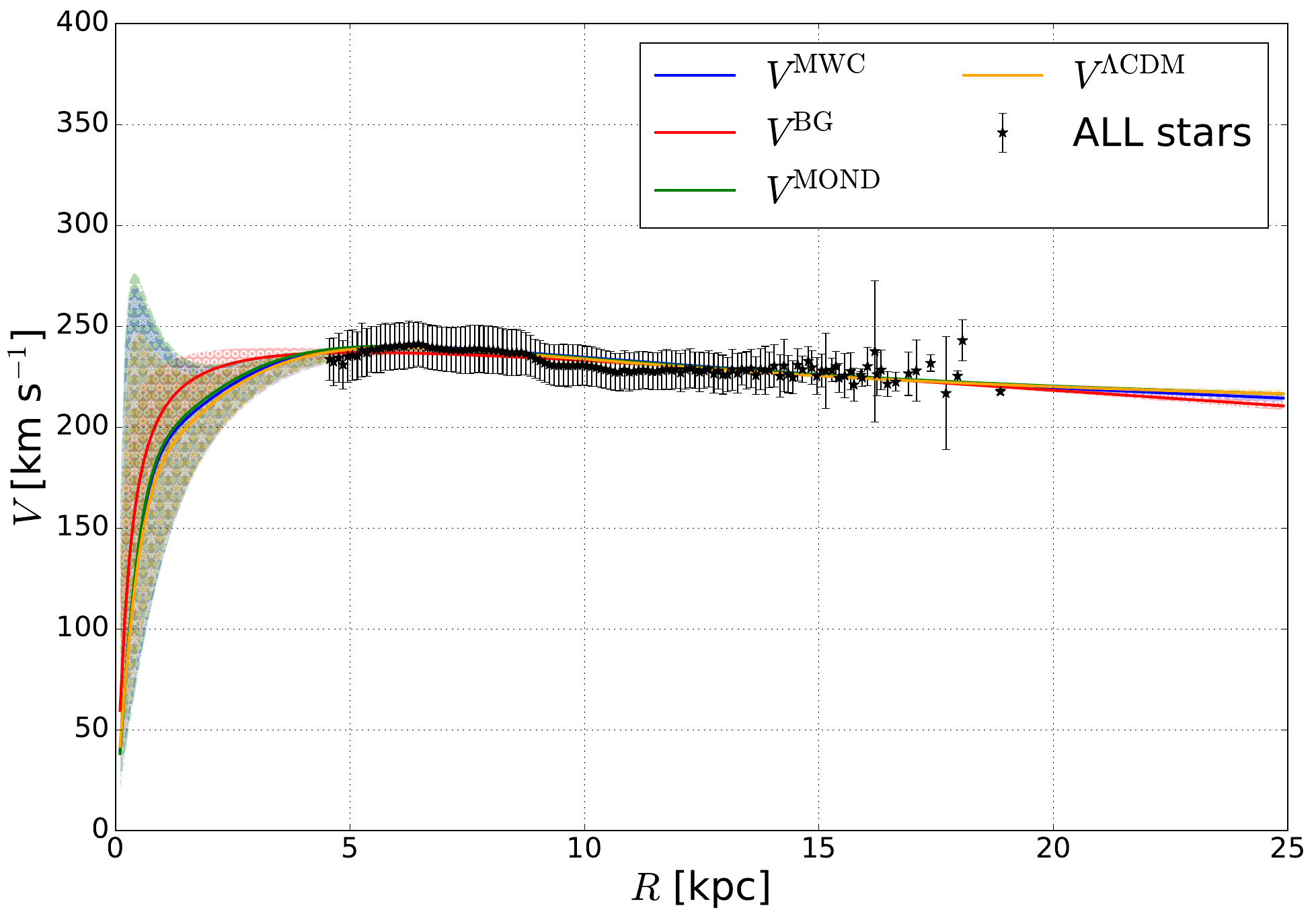}
    \end{subfigure}
    \caption{The azimuthal velocity profile of the MW as derived from the sample of disc tracers selected from Gaia DR3. For each dataset, the black starred symbols represent the median azimuthal velocity at the median distance from the galactic centre of the stellar population within each of the radial bins \cite[see tables A.1--A.6 of][]{beordoGeometrydrivenDarkmattersustainedMilky2024e}, where the RSE of the velocity distribution defines the corresponding error bar. The red, blue, green, and yellow curves show the best-fitting to the BG, MWC, MOND, and $\Lambda$CDM models, respectively. The filled areas represent the 68 per cent reliability intervals of each rotation curve; note that for $R \lesssim 4.5$ kpc both the classical and the relativistic curves are very uncertain because of the lack of data in that region.\label{fig:Vrot}}
\end{figure}

\begin{table}[htbp]
\centering
{\renewcommand{\arraystretch}{1.2}%
\resizebox{\textwidth}{!}{
\begin{tabular}{l|rrrrrr}
    \hline
    \textbf{MOND model} & OBA & DCEP & RGB & OBA $+$ DCEP & RGB $+$ DCEP & ALL \\
    \hline
    $M_{\rm b}$ [$10^{10}$ M$_{\odot}$] & $1.1^{+0.6}_{-0.6}$ & $1.2^{+0.6}_{-0.6}$ & $1.3^{+0.6}_{-0.6}$ & $1.1^{+0.6}_{-0.6}$ & $1.4^{+0.6}_{-0.6}$ & $1.3^{+0.6}_{-0.6}$ \\
    $b_{\rm b}$ [kpc] & $1.0^{+0.9}_{-0.6}$ & $0.8^{+0.8}_{-0.6}$ & $0.8^{+0.8}_{-0.5}$ & $0.9^{+0.9}_{-0.6}$ & $0.8^{+0.7}_{-0.5}$ & $0.8^{+0.8}_{-0.5}$ \\
    $M_{\rm td}$ [$10^{10}$ M$_{\odot}$] & $4.3^{+0.6}_{-0.7}$ & $4.3^{+0.7}_{-0.7}$ & $3.8^{+0.6}_{-0.6}$ & $4.1^{+0.7}_{-0.7}$ & $4.0^{+0.6}_{-0.6}$ & $4.0^{+0.6}_{-0.6}$ \\
    $a_{\rm td}$ [kpc] & $5.0^{+0.9}_{-0.8}$ & $5.2^{+0.9}_{-0.8}$ & $4.7^{+0.9}_{-0.9}$ & $4.4^{+0.7}_{-0.6}$ & $4.9^{+0.9}_{-0.8}$ & $4.8^{+0.8}_{-0.8}$ \\
    $b_{\rm td}$ [kpc] & $0.3^{+0.4}_{-0.1}$ & $0.4^{+0.5}_{-0.1}$ & $0.4^{+0.5}_{-0.1}$ & $0.3^{+0.3}_{-0.1}$ & $0.4^{+0.5}_{-0.1}$ & $0.3^{+0.5}_{-0.1}$ \\
    $M_{\rm Td}$ [$10^{10}$ M$_{\odot}$] & $4.7^{+0.7}_{-0.7}$ & $4.5^{+0.7}_{-0.8}$ & $4.2^{+0.7}_{-0.8}$ & $4.4^{+0.7}_{-0.7}$ & $4.2^{+0.7}_{-0.7}$ & $4.3^{+0.7}_{-0.8}$ \\
    $a_{\rm Td}$ [kpc] & $3.0^{+0.6}_{-0.6}$ & $2.9^{+0.7}_{-0.7}$ & $2.4^{+0.6}_{-0.6}$ & $2.8^{+0.6}_{-0.6}$ & $2.5^{+0.6}_{-0.6}$ & $2.5^{+0.6}_{-0.6}$ \\
    $b_{\rm Td}$ [kpc] & $0.9^{+0.8}_{-0.5}$ & $0.8^{+0.9}_{-0.5}$ & $0.4^{+0.7}_{-0.2}$ & $0.7^{+0.7}_{-0.4}$ & $0.5^{+0.7}_{-0.3}$ & $0.5^{+0.7}_{-0.3}$ \\
    $g_0$ [$10^{-10}$ m s$^{-2}$] & $1.20^{+0.02}_{-0.02}$ & $1.20^{+0.02}_{-0.02}$ & $1.20^{+0.02}_{-0.02}$ & $1.20^{+0.02}_{-0.02}$ & $1.20^{+0.02}_{-0.02}$ & $1.20^{+0.02}_{-0.02}$ \\
    WAIC & $-341\pm3$ & $-103\pm3$ & $-345\pm2$ & $-452\pm5$ & $-423\pm5$ & $-426\pm5$ \\
    LOO & $-341\pm3$ & $-103\pm3$ & $-345\pm2$ & $-452\pm5$ & $-423\pm5$ & $-426\pm5$ \\
    \hline
    \textbf{$\Lambda$CDM model} &  &  &  &  &  & \\
    \hline
    $M_{\rm b}$ [$10^{10}$ M$_{\odot}$] & $0.8^{+0.6}_{-0.5}$ & $0.9^{+0.6}_{-0.5}$ & $1.1^{+0.6}_{-0.6}$ & $0.8^{+0.6}_{-0.5}$ & $1.1^{+0.6}_{-0.6}$ & $1.0^{+0.6}_{-0.6}$ \\
    $b_{\rm b}$ [kpc] & $0.9^{+0.8}_{-0.6}$ & $0.8^{+0.8}_{-0.6}$ & $0.8^{+0.8}_{-0.5}$ & $0.9^{+0.9}_{-0.6}$ & $0.8^{+0.8}_{-0.5}$ & $0.8^{+0.8}_{-0.5}$ \\
    $M_{\rm td}$ [$10^{10}$ M$_{\odot}$] & $3.8^{+0.7}_{-0.7}$ & $3.7^{+0.8}_{-0.8}$ & $3.6^{+0.7}_{-0.7}$ & $4.1^{+0.7}_{-0.7}$ & $3.7^{+0.7}_{-0.7}$ & $3.7^{+0.7}_{-0.7}$ \\
    $a_{\rm td}$ [kpc] & $4.8^{+0.9}_{-0.8}$ & $5.1^{+0.9}_{-0.9}$ & $4.7^{+1.0}_{-0.9}$ & $5.0^{+0.8}_{-0.7}$ & $5.0^{+1.0}_{-0.9}$ & $4.9^{+0.9}_{-0.9}$ \\
    $b_{\rm td}$ [kpc] & $0.3^{+0.4}_{-0.1}$ & $0.3^{+0.5}_{-0.1}$ & $0.3^{+0.5}_{-0.1}$ & $0.3^{+0.3}_{-0.1}$ & $0.3^{+0.6}_{-0.1}$ & $0.3^{+0.5}_{-0.1}$ \\
    $M_{\rm Td}$ [$10^{10}$ M$_{\odot}$] & $3.9^{+0.9}_{-0.9}$ & $3.7^{+0.9}_{-0.9}$ & $3.8^{+0.9}_{-0.9}$ & $4.2^{+0.8}_{-0.8}$ & $3.7^{+0.8}_{-0.8}$ & $3.8^{+0.8}_{-0.8}$ \\
    $a_{\rm Td}$ [kpc] & $2.9^{+0.6}_{-0.7}$ & $2.8^{+0.7}_{-0.7}$ & $2.5^{+0.6}_{-0.6}$ & $3.1^{+0.6}_{-0.7}$ & $2.6^{+0.6}_{-0.6}$ & $2.7^{+0.6}_{-0.6}$ \\
    $b_{\rm Td}$ [kpc] & $0.8^{+0.8}_{-0.5}$ & $0.7^{+0.9}_{-0.5}$ & $0.5^{+0.7}_{-0.3}$ & $1.0^{+0.9}_{-0.6}$ & $0.5^{+0.8}_{-0.3}$ & $0.6^{+0.8}_{-0.3}$ \\
    $\log(\rho_{\rm s} / [$M$_{\odot}$ kpc$^{-3}])$ & $5.8^{+0.3}_{-0.3}$ & $5.9^{+0.3}_{-0.3}$ & $5.8^{+0.3}_{-0.3}$ & $6.0^{+0.3}_{-0.3}$ & $5.9^{+0.3}_{-0.3}$ & $5.9^{+0.3}_{-0.3}$ \\
    $r_{\rm s}$ [kpc] & $38^{+20}_{-12}$ & $35^{+16}_{-10}$ & $35^{+17}_{-11}$ & $25^{+10}_{-7}$ & $29^{+11}_{-8}$ & $29^{+11}_{-8}$ \\
    $\alpha$ & $0.15^{+0.06}_{-0.04}$ & $0.14^{+0.05}_{-0.04}$ & $0.14^{+0.05}_{-0.04}$ & $0.11^{+0.04}_{-0.03}$ & $0.12^{+0.04}_{-0.03}$ & $0.12^{+0.04}_{-0.03}$ \\
    $V_{200}$ [km s$^{-1}$] & $192^{+27}_{-20}$ & $186^{+22}_{-18}$ & $179^{+20}_{-16}$ & $164^{+12}_{-13}$ & $174^{+13}_{-14}$ & $173^{+13}_{-14}$ \\
    $C_{200}$ & $7.5^{+2.4}_{-1.9}$ & $8.0^{+2.5}_{-2.0}$ & $7.7^{+2.5}_{-1.9}$ & $9.8^{+3.0}_{-2.3}$ & $8.8^{+2.5}_{-2.0}$ & $8.8^{+2.6}_{-2.0}$ \\
    $\log(M_{200} / [$M$_{\odot}])$ & $12.39^{+0.17}_{-0.14}$ & $12.35^{+0.15}_{-0.13}$ & $12.30^{+0.14}_{-0.12}$ & $12.19^{+0.09}_{-0.10}$ & $12.26^{+0.10}_{-0.11}$ & $12.26^{+0.09}_{-0.11}$ \\
    $r_{200}$ [kpc] & $286^{+40}_{-30}$ & $278^{+33}_{-27}$ & $267^{+30}_{-24}$ & $245^{+17}_{-19}$ & $259^{+20}_{-21}$ & $258^{+19}_{-20}$ \\
    WAIC & $-341\pm4$ & $-103\pm3$ & $-345\pm2$ & $-450\pm5$ & $-425\pm6$ & $-427\pm6$ \\
    LOO & $-341\pm4$ & $-103\pm3$ & $-345\pm2$ & $-450\pm5$ & $-425\pm7$ & $-428\pm7$ \\
    \hline
\end{tabular}}}
\caption{Top: estimates of the free parameters of the MOND model, namely, the medians of the posterior distributions of the Bayesian analysis for each dataset; the upper and lower bounds (estimated with the $15.87^{\rm th}$ and $84.13^{\rm th}$ percentiles) enclose their corresponding $ 1\sigma $ credible intervals.
    $M_{\rm b}$ and $b_{\rm b}$ are, respectively, the mass and the Plummer radius of the bulge; $M_{\rm td}$, $a_{\rm td}$, and $b_{\rm td}$ are the mass, scale length, and scale height of the thin disc; $M_{\rm Td}$, $a_{\rm Td}$, and $b_{\rm Td}$ are the mass, scale length, and scale height of the thick disc; $g_0$ is the Mondian acceleration scale. 
    Bottom: estimates of the free parameters for the $\Lambda$CDM model: in addition the parameters of the baryonic matter distribution, $\rho_{\rm s}$, $r_s$, and $\alpha$, are the three parameters defining the Einasto profile, i.e., the characteristic density, scale radius, and shape parameter; $V_{200}$, $C_{200}$, $M_{200}$ are, respectively, the rotation velocity, concentration, and halo mass calculated at the virial radius $r_{200}$. Log-values of the WAIC and LOO tests for the Bayesian model comparison are also reported. \label{table:fit_params}}
\end{table}
The parameters of the baryonic matter distribution are found in agreement between different datasets and models, even though the $\Lambda$CDM paradigm tends to assign less mass to the baryonic component, as the values of $M_{\rm b}$, $M_{\rm td}$, $M_{\rm Td}$ are slightly smaller compared to those estimated with the MOND and MWC models.
Additionally, $\Lambda$CDM cosmological constraints are observed, as made clear by figure~\ref{fig:LCDMpriors}: specifically, the estimated parameters align within 1-$\sigma$ with the relations for the halo mass-concentration and the Einasto shape parameter, and within 2-$\sigma$ with the SHM relation.
Rotation curves of the Milky Way for six stellar populations are presented in figure~\ref{fig:Vrot}: the results for MOND and $\Lambda$CDM are drawn on top of the corresponding results of \cite{beordoGeometrydrivenDarkmattersustainedMilky2024e} for the BG and MWC models.
Again, the fits of the four models result statistically equivalent to each other, having comparable values for the WAIC and LOO tests.
%
% This result might suggest a potential issue of data over-fitting.
% To reduce the possible over-fitting, a more rigorous approach could be pursued, but, keeping in mind the significant amount of data, we leave this to future work.

\begin{figure}[htbp]
   %\centering % Not needed
   \begin{subfigure}{.495\linewidth}
        \includegraphics[width=\linewidth]{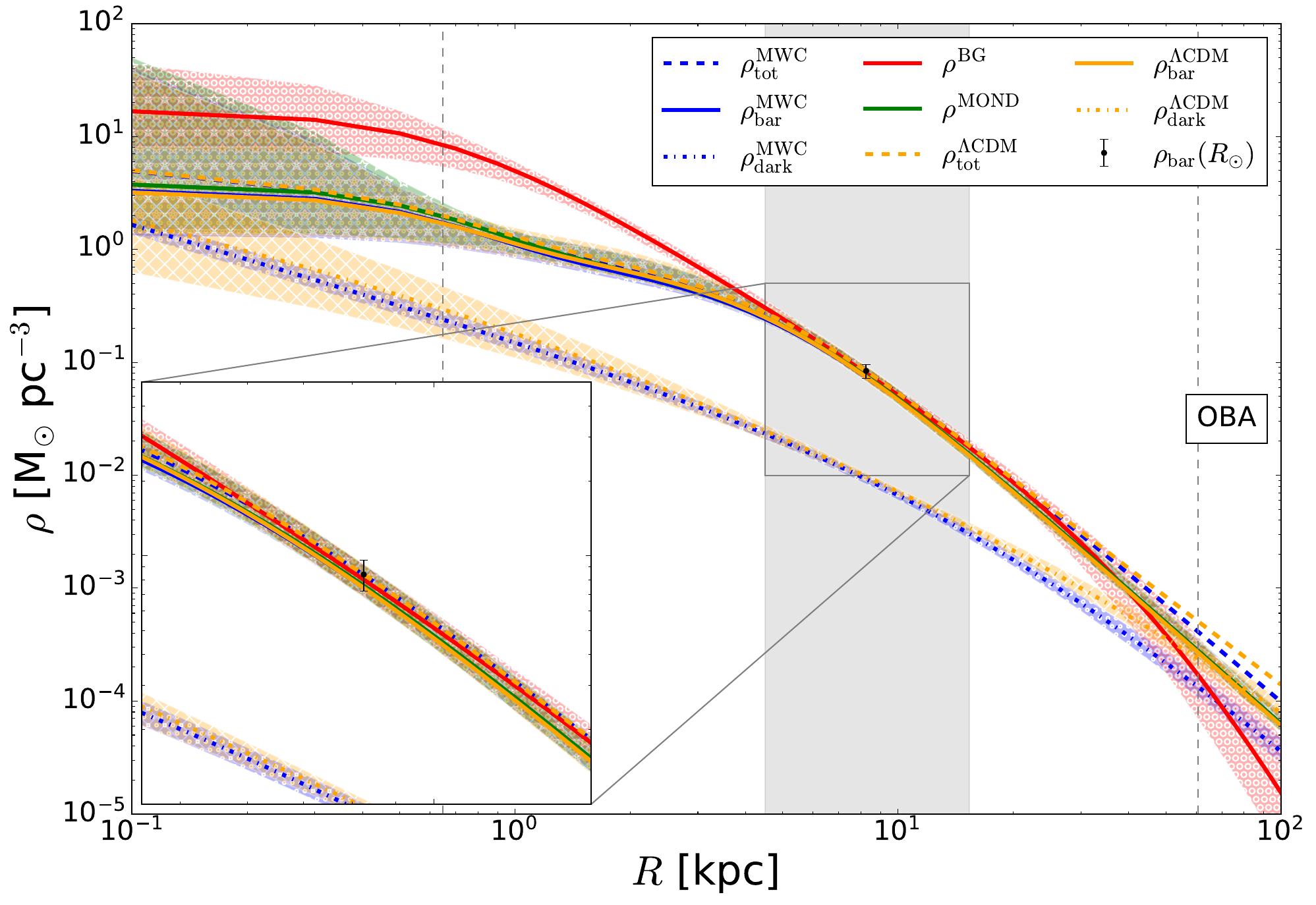}
    \end{subfigure}
    \hfill
    \begin{subfigure}{.495\linewidth}
        \includegraphics[width=\linewidth]{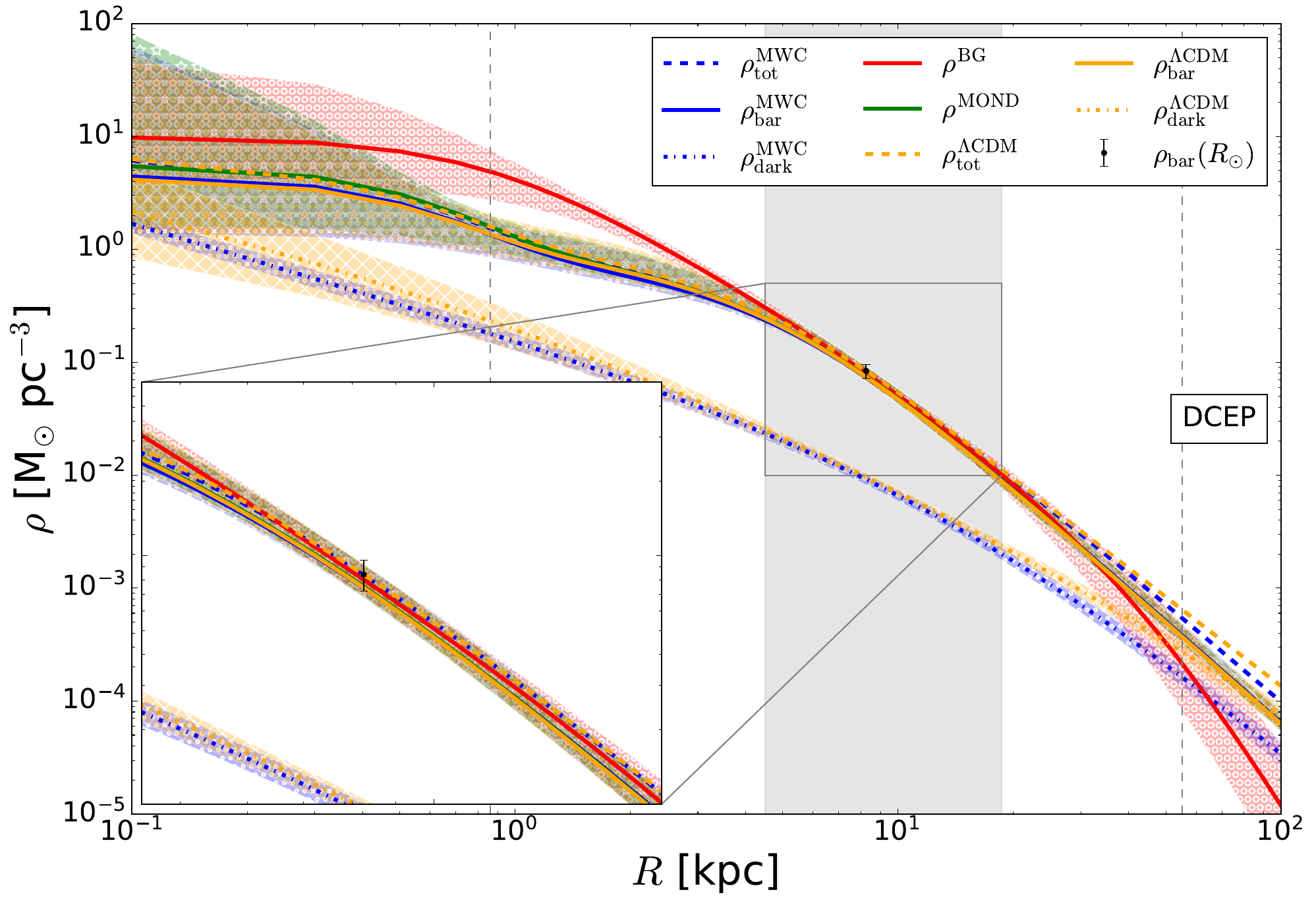}
    \end{subfigure}
    %% leave a blank line to create a line break

    \begin{subfigure}{.495\linewidth}
        \includegraphics[width=\linewidth]{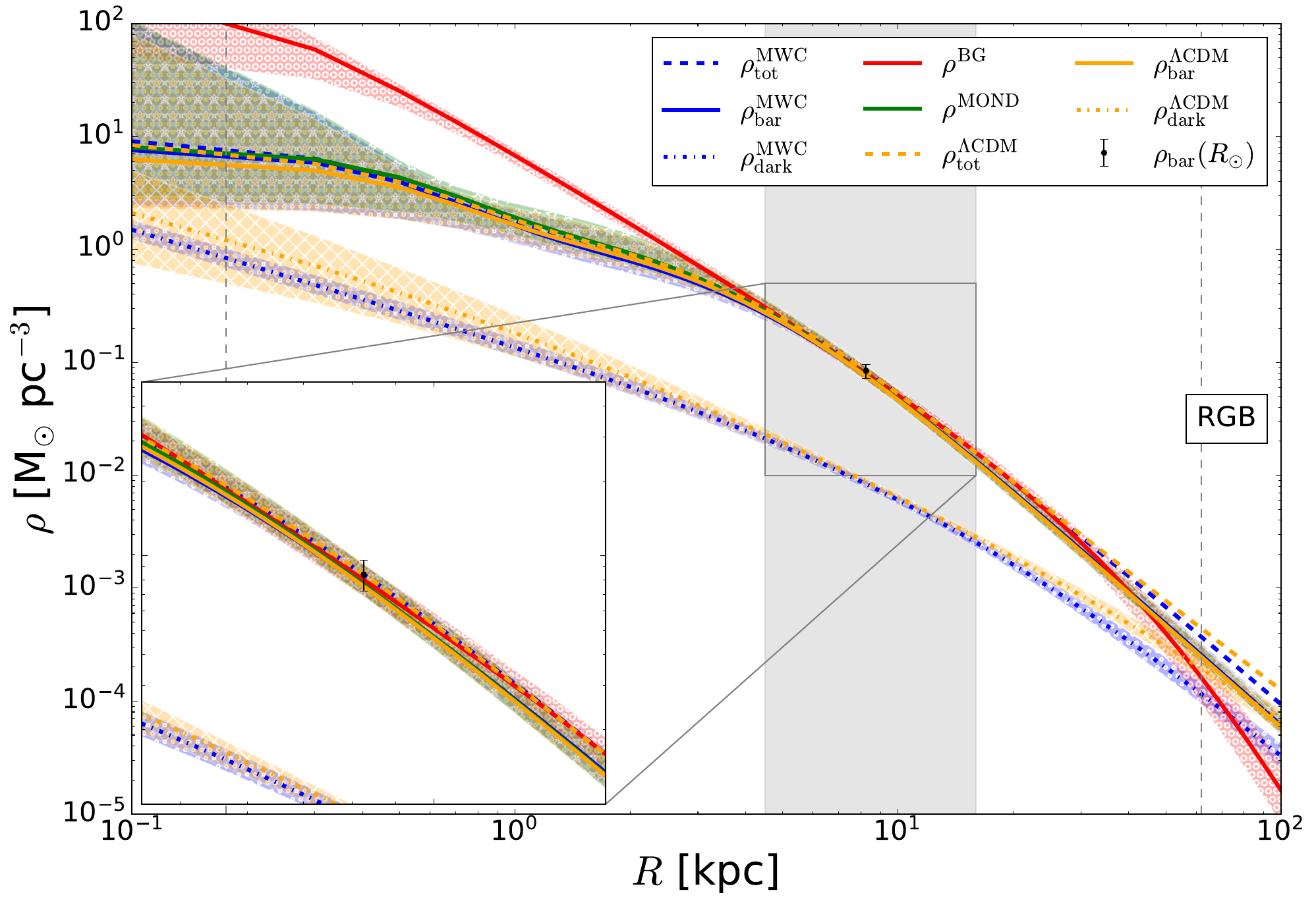}
    \end{subfigure}
    \hfill
    \begin{subfigure}{.495\linewidth}
        \includegraphics[width=\linewidth]{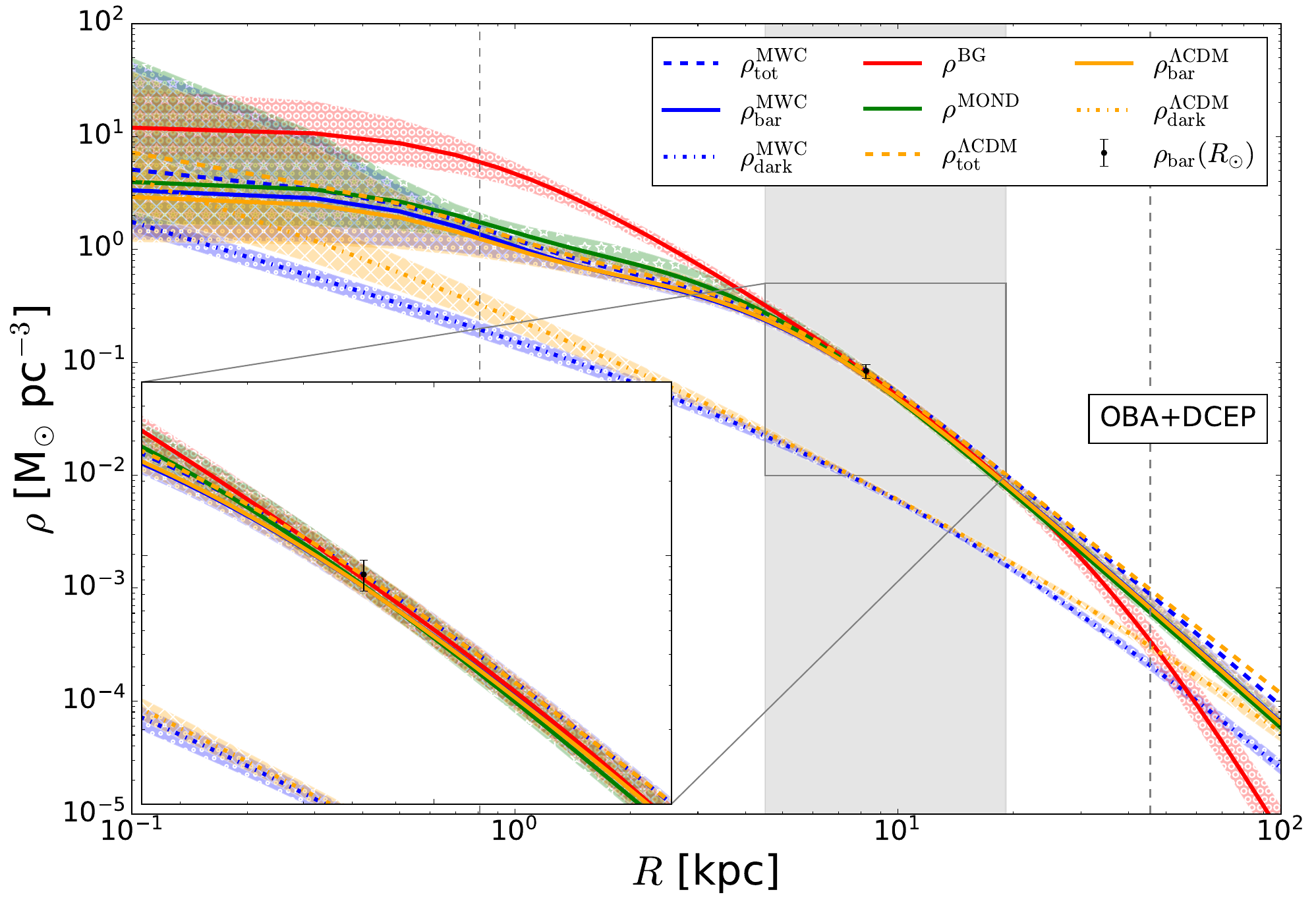}
    \end{subfigure}
     %% leave a blank line to create a line break

    \begin{subfigure}{.495\linewidth}
        \includegraphics[width=\linewidth]{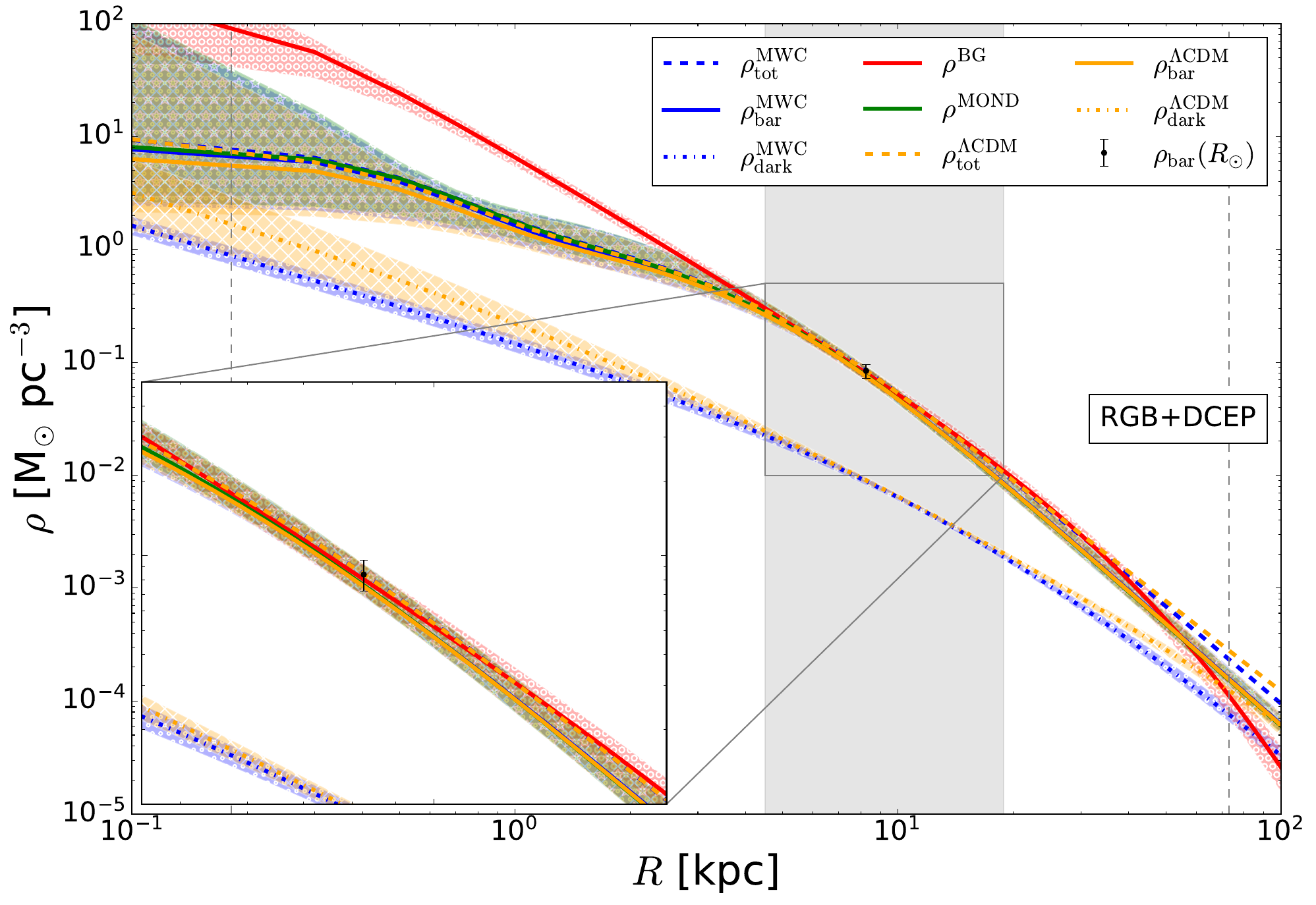}
    \end{subfigure}
    \hfill
    \begin{subfigure}{.495\linewidth}
        \includegraphics[width=\linewidth]{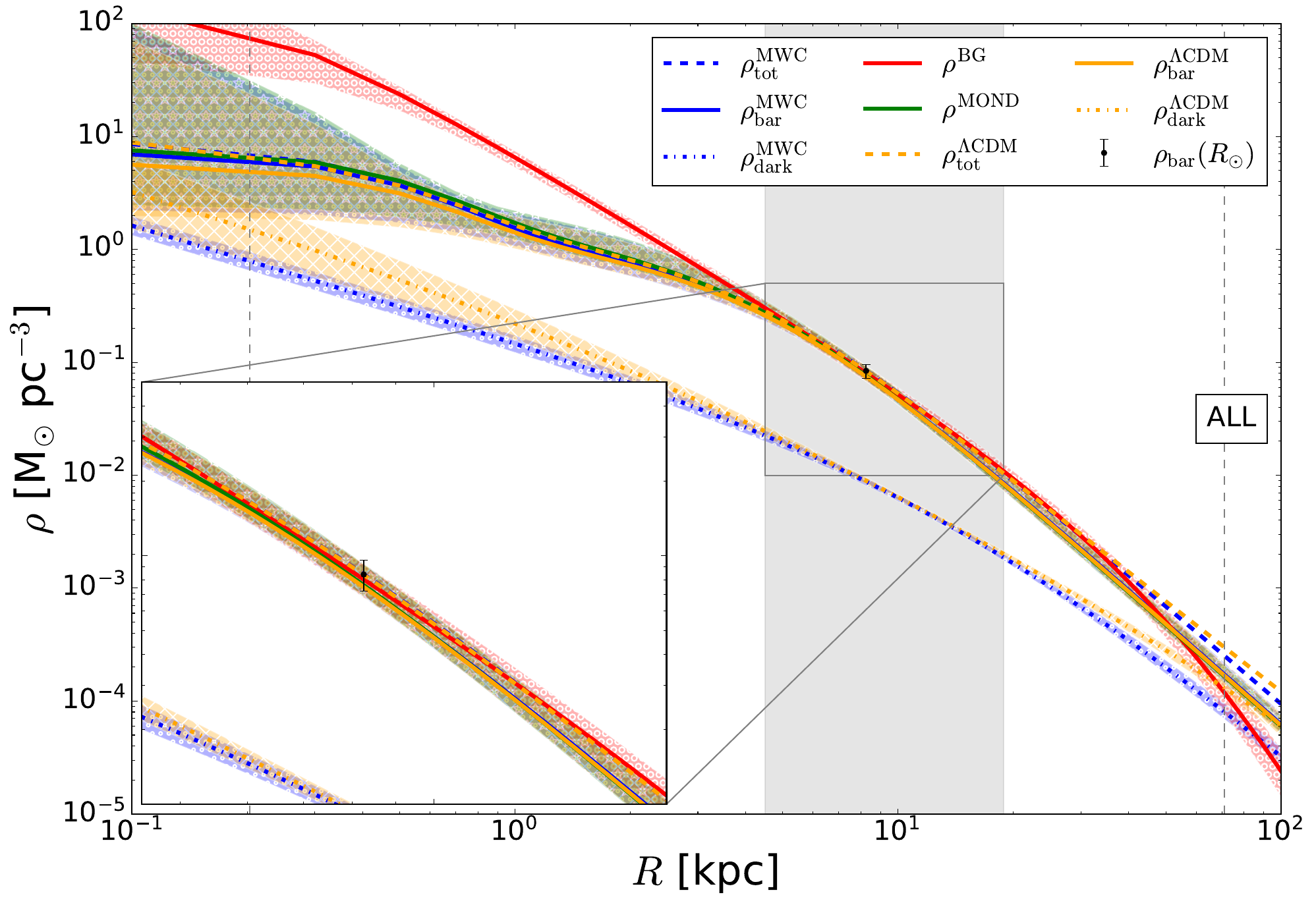}
    \end{subfigure}
    \caption{Density profiles of the MW at $z=0$ for the four models, with their corresponding 68 per cent confidence intervals; in each panel, the red, blue, green, and orange solid lines represent the baryonic matter contributions for the BG, MWC, MOND, and $\Lambda$CDM models respectively. The blue and orange dashed lines show the total matter for the MWC and $\Lambda$CDM models, while the dash-dotted lines are the corresponding dark matter contributions. The vertical grey dashed lines represent the values of $r_{\rm in}$ and $R_{\rm out}$ of the BG model, while the vertical grey band spans the radial range covered by the sample. Finally, the black dot represents the local mass density inferred at the Sun position, i.e. $\rho_{\rm bar}(R_{\odot}) = 0.084 \pm 0.012$ M$_\odot {\rm pc}^{-3}$ from \cite{mckeeSTARSGASDARK2015}.}
    \label{fig:dens}
\end{figure}

Figure~\ref{fig:dens} shows the matter density profiles for the four models along with the only data point considered, which corresponds to the baryonic matter density observed at the Sun \cite[$\rho_{\rm bar}(R_{\odot}) = 0.084 \pm 0.012$ M$_\odot {\rm pc}^{-3}$,][]{mckeeSTARSGASDARK2015}.
Estimates of the local baryonic density ($\rho^{\rm \Lambda CDM}_{\rm bar, \odot}$ and $\rho^{\rm MOND}_{\odot}$, table~\ref{table:massdens}) are consistent with those in \cite{beordoGeometrydrivenDarkmattersustainedMilky2024e}, being all around $0.080\, {\rm M}_{\odot} {\rm pc}^{-3}$, as well as with estimates found in the literature \cite{Garbari2012, Bienayme2014, mckeeSTARSGASDARK2015}.
As for the dark matter density at the Sun ($\rho^{\rm \Lambda CDM}_{\rm h, \odot}$), we recover again the recent values reported in the literature \cite{eilersCircularVelocityCurve2019, Cautun2020, widmarkWeighingGalacticDisk2021, wangMilkyWayTotal2022}, being almost ten times smaller than the local baryonic density: for the full sample, $\rho^{\rm \Lambda CDM}_{\rm h, \odot}= (0.0090 \pm 0.0006)$ M$_\odot {\rm pc}^{-3}$, which corresponds to $(0.34 \pm 0.02)$ GeV cm$^{-3}$. 

\begin{table}[htbp]
\centering
{\renewcommand{\arraystretch}{1.2}%
\resizebox{\textwidth}{!}{
\begin{tabular}{l|rrrrrr}
    \hline
    Quantity & OBA & DCEP & RGB & OBA $+$ DCEP & RGB $+$ DCEP & ALL \\
    \hline
    $\rho^{\Lambda \rm CDM}_{\rm bar, \odot}$ [${\rm M}_{\odot}{\rm pc}^{-3}$] &  $0.080^{+0.012}_{-0.012}$ & $0.080^{+0.012}_{-0.012}$ &    $0.080^{+0.012}_{-0.012}$ & $0.080^{+0.012}_{-0.012}$ &  $0.080^{+0.012}_{-0.012}$ & $0.080^{+0.012}_{-0.012}$ \\
    $\rho^{\Lambda \rm CDM}_{\rm h, \odot}$ [${\rm M}_{\odot}{\rm pc}^{-3}$] &  $0.0099^{+0.0008}_{-0.0008}$ & $0.0098^{+0.0008}_{-0.0008}$ &    $0.0088^{+0.0078}_{-0.0007}$ & $0.0085^{+0.0006}_{-0.0006}$ &  $0.0091^{+0.0006}_{-0.0006}$ & $0.0090^{+0.0006}_{-0.0006}$ \\
    $\rho^{\rm MOND}_{\odot}$ [${\rm M}_{\odot}{\rm pc}^{-3}$] & $0.080^{+0.012}_{-0.012}$ & $0.080^{+0.012}_{-0.012}$ & $0.081^{+0.012}_{-0.012}$ & $0.081^{+0.012}_{-0.012}$ & $0.081^{+0.012}_{-0.012}$ & $0.081^{+0.012}_{-0.012}$ \\
    $M^{\Lambda \rm CDM}_{\rm bar}$ [$10^{10} \, {\rm M}_{\odot}$] &  $8.49^{+1.10}_{-1.03}$ &  $8.31^{+1.11}_{-1.04}$ &  $8.51^{+1.06}_{-0.95}$ &  $9.23^{+0.95}_{-0.93}$ &  $8.43^{+0.96}_{-0.91}$ &  $8.49^{+0.96}_{-0.89}$ \\
    $M^{\rm MOND}$ [$10^{10} \, {\rm M}_{\odot}$] &  $10.12^{+0.33}_{-0.30}$ &  $10.05^{+0.48}_{-0.45}$ &  $9.32^{+0.27}_{-0.24}$ &  $9.59^{+0.19}_{-0.18}$ &  $9.59^{+0.21}_{-0.19}$ &  $9.56^{+0.21}_{-0.19}$ \\
    $M^{\Lambda \rm CDM}_{\rm 200}$ [$10^{12} \, {\rm M}_{\odot}$] & $2.45^{+1.16}_{-0.68}$ & $2.24^{+0.88}_{-0.59}$ & $1.99^{+0.74}_{-0.49}$ & $1.53^{+0.35}_{-0.33}$ & $1.82^{+0.44}_{-0.40}$ & $1.80^{+0.43}_{-0.39}$\\
    \hline
    \end{tabular}}}
    \caption{Estimates of the local baryonic mass density $\rho_{\rm bar}(R_{\odot})$ for each dataset and both models (respectively $\rho^{\rm \Lambda CDM}_{\rm bar, \odot}$ and $\rho^{\rm MOND}_{\odot}$) and of the local dark matter density $\rho^{\rm \Lambda CDM}_{\rm h, \odot}$; the upper and lower bounds (estimated with the $15.87^{\rm th}$ and $84.13^{\rm th}$ percentiles) enclose their corresponding $ 1\sigma $ credible intervals. $M^{\rm \Lambda CDM}_{\rm bar}$ and $M^{\rm MOND}$ are, respectively, the baryonic mass for the $\Lambda$CDM model and the total mass for the MOND model; while $M^{\rm \Lambda CDM}_{\rm 200}$ is the estimated virial mass of the Milky Way in the $\Lambda$CDM scenario. For a direct comparison with the BG and MWC models, similar estimates can be found in table 2 of \cite{beordoGeometrydrivenDarkmattersustainedMilky2024e}. \label{table:massdens}}
\end{table}
As expected, all density profiles show agreement within the radial range covered by the data: MWC and $\Lambda$CDM total matter density profiles are almost coincident, while departing from each other only at very large radii; so do their dark matter density profiles, but the Einasto profiles of the $\Lambda$CDM model result larger than the NFW ones both in the inner and outer parts of the Galaxy; this translates into the dynamical mass being supplied by more dark matter in the $\Lambda$CDM scenario compared to the case of an NFW halo without cosmological constraints. 

This is confirmed by table~\ref{table:massdens}, where the estimates for the total baryonic and virial masses are reported (respectively $M^{\Lambda \rm CDM}_{\rm bar}$ and $M^{\Lambda \rm CDM}_{\rm 200}$). The baryonic mass predicted by the $\Lambda$CDM scenario is around $8\text{--}9 \times 10^{10} \, {\rm M}_{\odot}$, smaller than the values of $9\text{--}10 \times 10^{10} \, {\rm M}_{\odot}$ expected for the MWC model \cite[see table 2 in][]{beordoGeometrydrivenDarkmattersustainedMilky2024e}, 
 %  (see table  2 in Beordo et al 2024) >>> Cite the proper table in Beordo et al  2024 come e' fatto nel testo della caption della table 2 di questo MOND 
but still higher compared to \cite{mcmillanMassDistributionGravitational2017}, which reports a value of $ (5.43 \pm 0.57) \times 10^{10} \, {\rm M}_{\odot} $. The virial mass, ranging in the interval $1.5\text{--}2.5 \times 10^{12} \, {\rm M}_{\odot}$, is up to two times the value of $ (1.30 \pm 0.30) \times 10^{12} \, {\rm M}_{\odot} $ found by \cite{mcmillanMassDistributionGravitational2017} and the values in \cite{beordoGeometrydrivenDarkmattersustainedMilky2024e} for the virial mass of the MWC model, significantly higher than those reported by some authors \cite[around $5\text{--}8 \times 10^{11} \, {\rm M}_{\odot}$]{labiniMassModelsMilky2023, eilersCircularVelocityCurve2019} and an order of magnitude greater than some very recent values claimed \cite[around $1\text{--}2 \times 10^{11} \, {\rm M}_{\odot}$]{jiaoDetectionKeplerianDecline2023, ouDarkMatterProfile2024}. These low values from the literature are the result of a clearly falling rotation curve derived by the authors beyond 18 kpc. The differences in the data selection and construction of the rotation curve, listed in section~\ref{sec:keplerian_fall}, are potential contributors to the observed discrepancy in the estimation of the virial mass compared to recent literature. As a result, our findings suggest that the dynamical mass of the Galaxy, in the classical context with dark matter, is expected to remain on the order of $10^{12} \, {\rm M}_{\odot}$.

The BG and MOND density profiles are consistent with both the baryonic and total density profiles of the classical models (MWC and $\Lambda$CDM). Therefore, no further insight can be gained at this stage without looking at the amount of mass expected.
The total mass predicted by MOND ($M^{\rm MOND}$) aligns with the total baryonic mass of the MWC model, while being up to 10 per cent higher than the $\Lambda$CDM estimate. Moreover, this value exceeds the estimate of $ (7.8 \pm 0.5) \times 10^{10} \, {\rm M}_{\odot} $ found by \cite{labiniMassModelsMilky2023} in the MOND framework.
Finally, the BG relativistic mass reported in \cite{beordoGeometrydrivenDarkmattersustainedMilky2024e} compares favourably with the value of $1\text{--}2 \times 10^{10} \, {\rm M}_{\odot}$ expected for the baryonic mass in $\Lambda$CDM and MOND, calculated in the region of validity of the BG model.

\subsection{Contributions to the rotation curve}
\begin{figure}[htbp]
   \begin{subfigure}{.495\linewidth}
        \includegraphics[width=\linewidth]{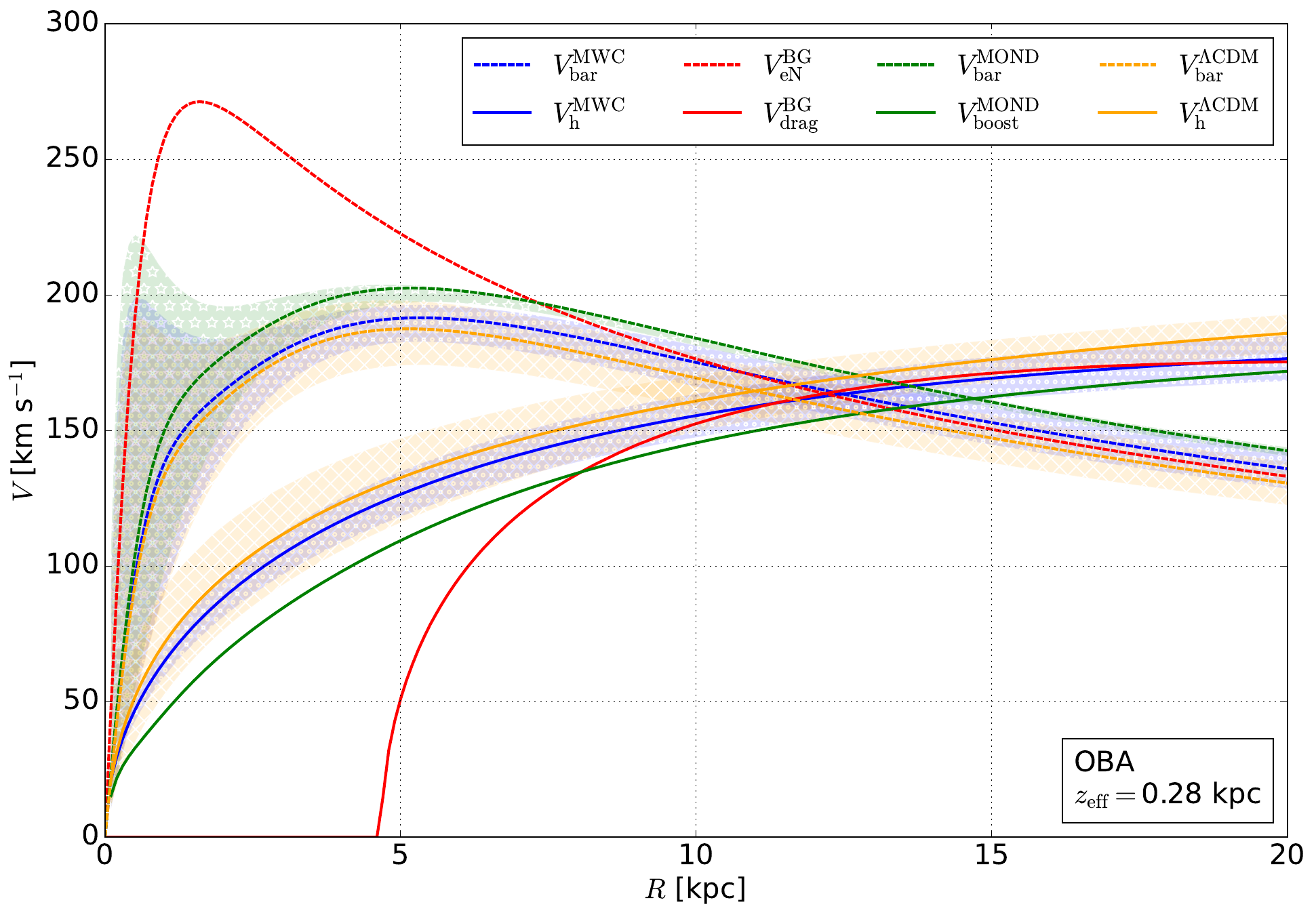}
    \end{subfigure}
    \hfill
    \begin{subfigure}{.495\linewidth}
        \includegraphics[width=\linewidth]{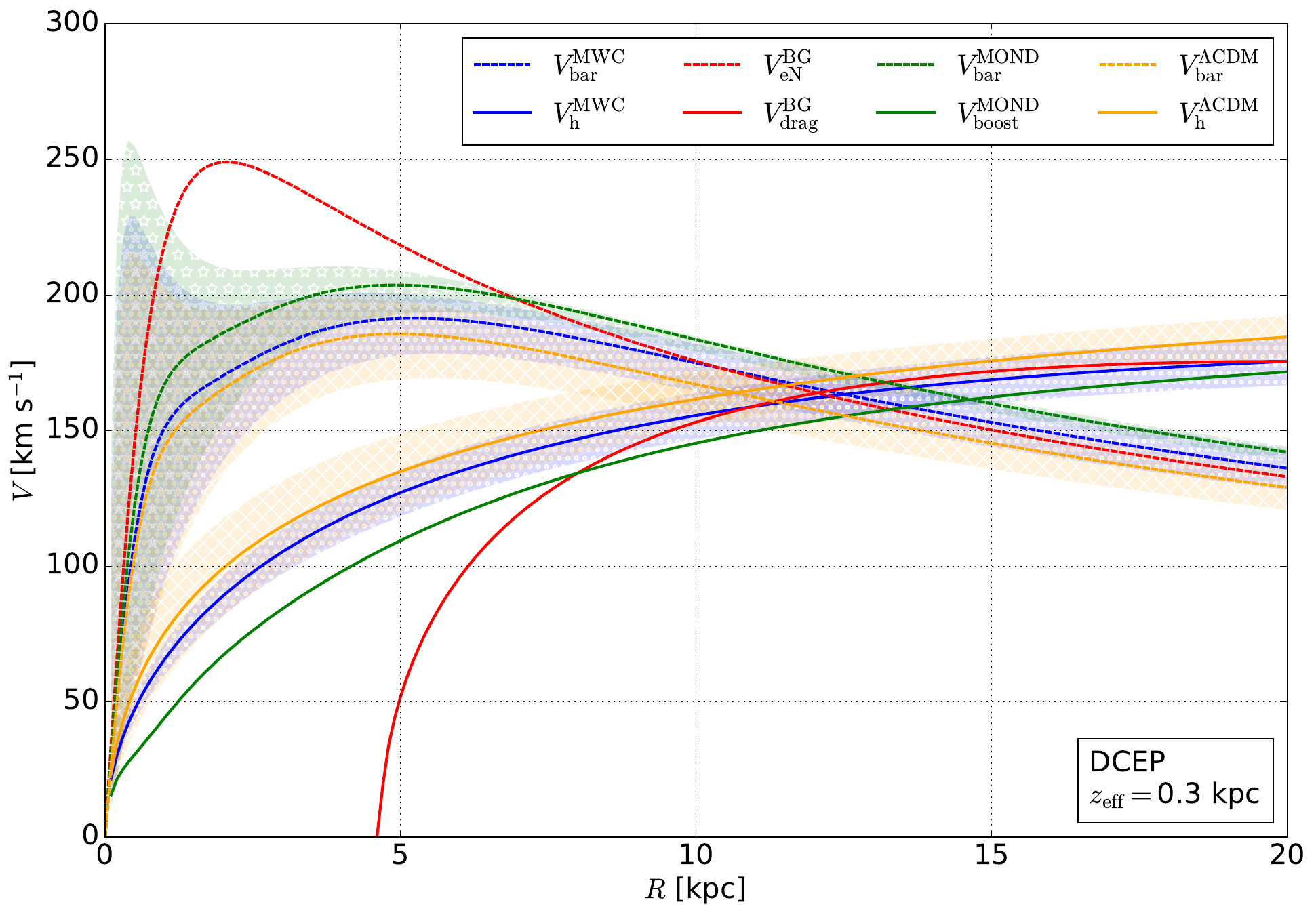}
    \end{subfigure}
    %% leave a blank line to create a line break

    \begin{subfigure}{.495\linewidth}
        \includegraphics[width=\linewidth]{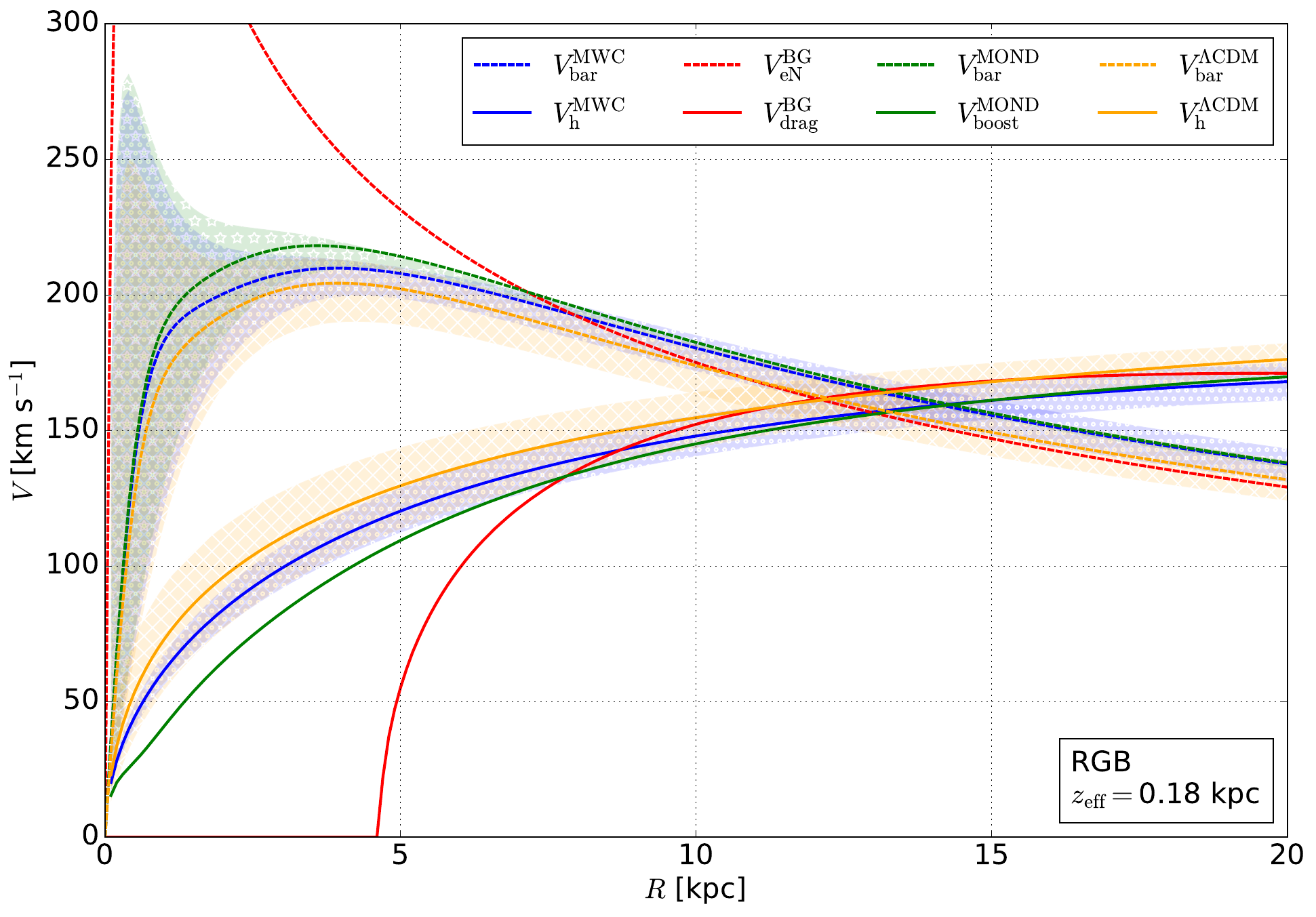}
    \end{subfigure}
    \hfill
    \begin{subfigure}{.495\linewidth}
        \includegraphics[width=\linewidth]{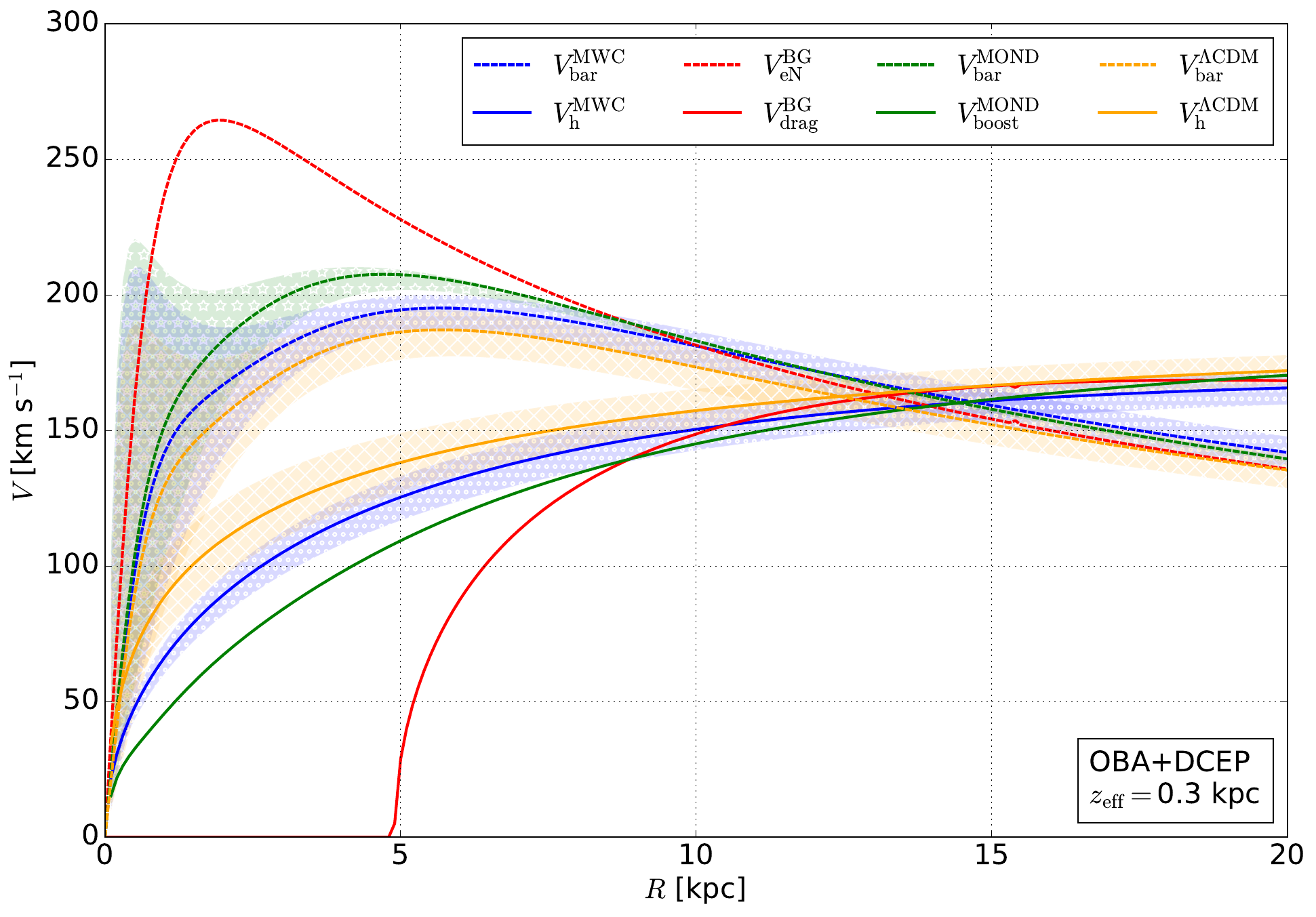}
    \end{subfigure}
     %% leave a blank line to create a line break

    \begin{subfigure}{.495\linewidth}
        \includegraphics[width=\linewidth]{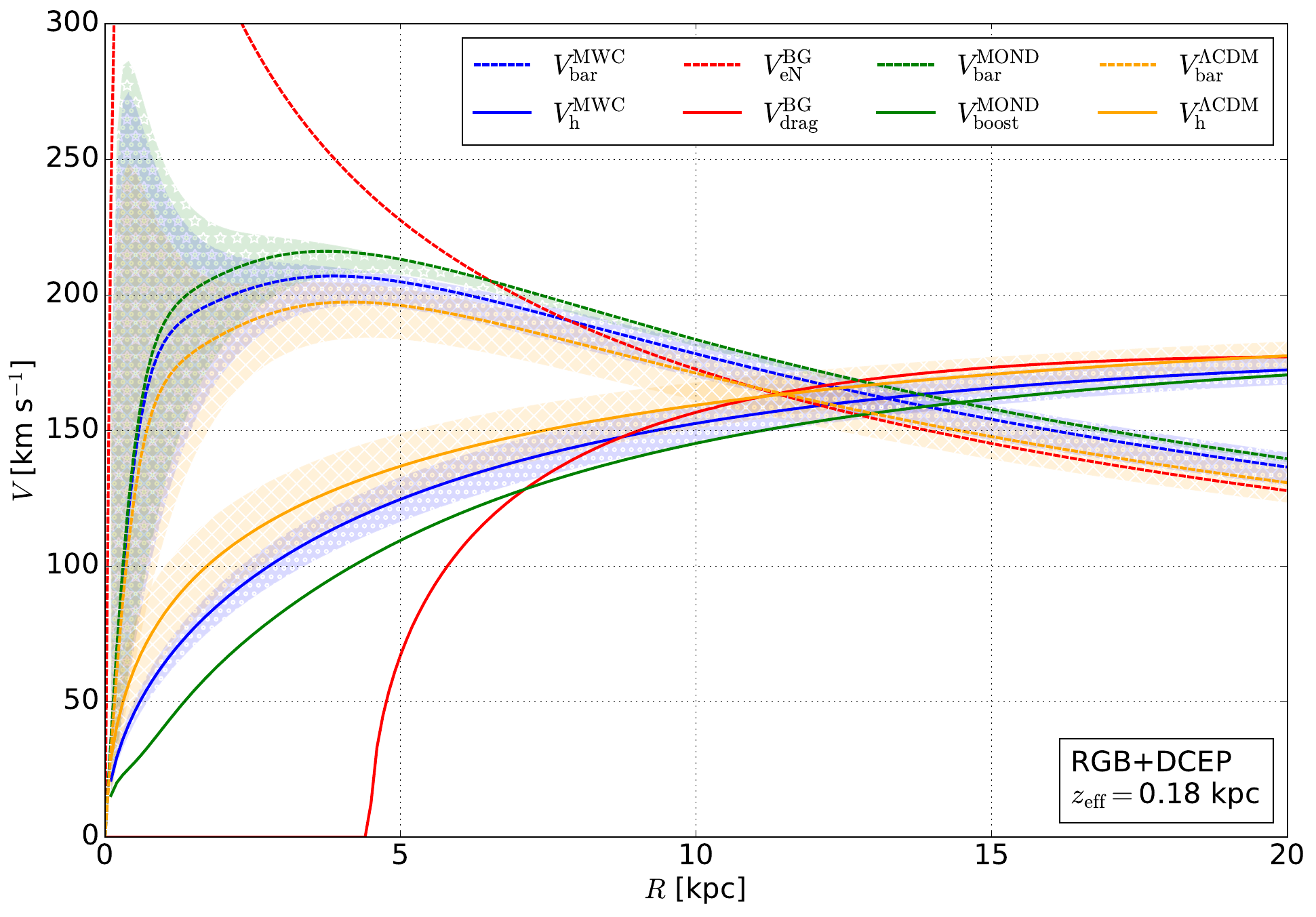}
    \end{subfigure}
    \hfill
    \begin{subfigure}{.495\linewidth}
        \includegraphics[width=\linewidth]{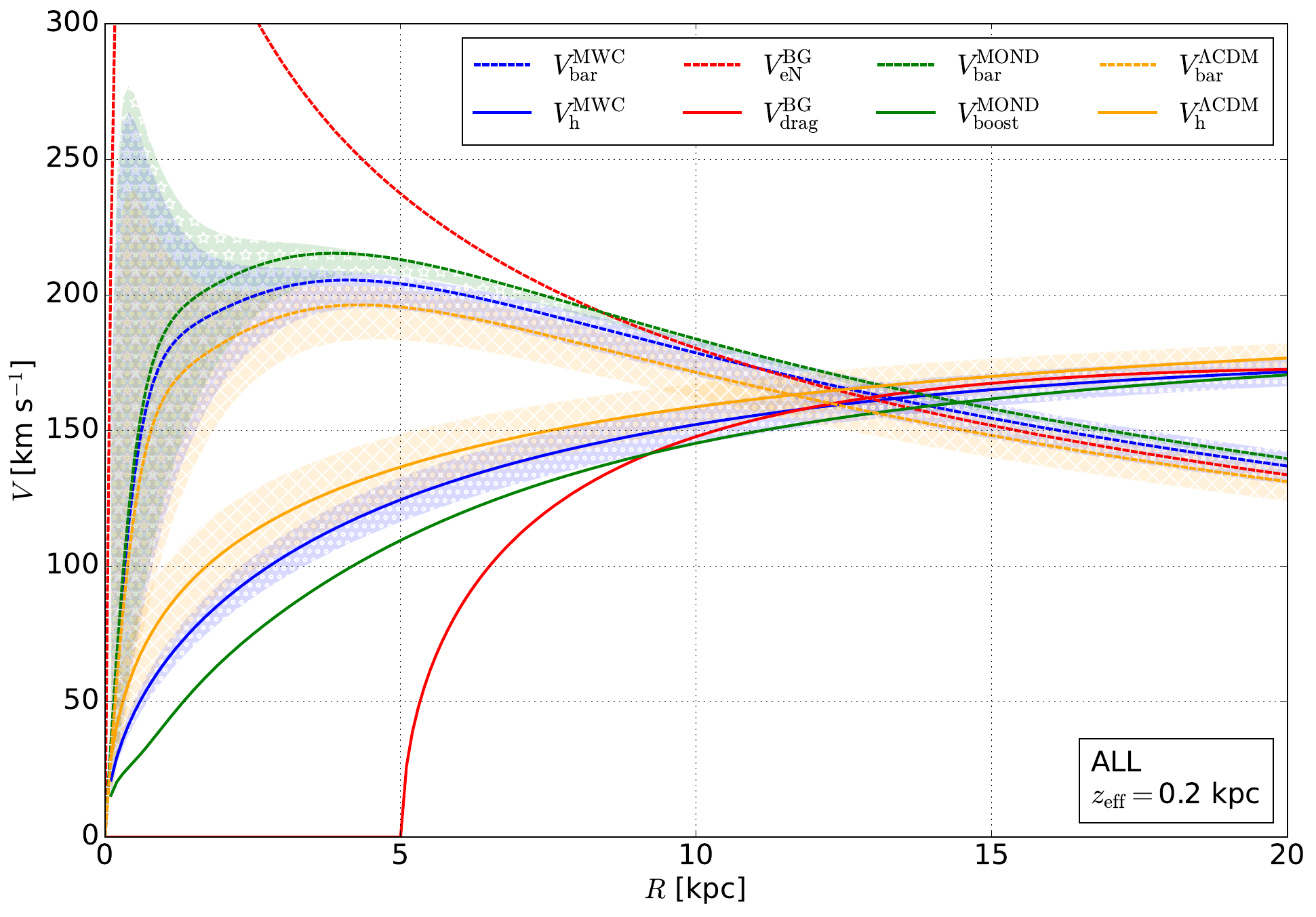}
    \end{subfigure}
    \caption{Red, blue, green and yellow lines refer to the BG, MWC, MOND, and $\Lambda$CDM models, respectively. The dashed lines represent the Newtonian/baryonic counterparts to the rotation curves: $V_{\rm eN}^{\rm BG}$ is the relativistic effective Newtonian velocity for the BG model, $V_{\rm bar}^{\rm MWC}$ and $V_{\rm bar}^{\rm \Lambda CDM}$ are the velocities contributed by the baryonic matter for the MWC and $\Lambda$CDM models, and $V_{\rm bar}^{\rm MOND}$ is the Newtonian velocity contributed by the baryonic matter for the MOND model. The solid lines show the MWC and $\Lambda$CDM halo components alone, respectively $V_{\rm h}^{\rm MWC}$ and $V_{\rm h}^{\rm \Lambda CDM}$, the gravitational dragging contribution for the BG model, i.e. $V_{\rm drag}^{\rm BG}$, and the Mondian boost $V_{\rm acc}^{\rm MOND}$. In the bottom right corner of each panel, $z_{\rm eff}$ represents the effective vertical width of validity for the relativistic disc in the BG framework.\label{fig:components}}
\end{figure}

We can compare the models by making the contributions to the rotation curves explicit, recalling what was done in section 7 of \cite{beordoGeometrydrivenDarkmattersustainedMilky2024e}, since different models and samples contribute to the rotation curve differently. Figure~\ref{fig:components}, drawn on top of figure 4 in \cite{beordoGeometrydrivenDarkmattersustainedMilky2024e}, shows the Newtonian/baryonic counterpart for each model, alongside the corresponding non-Newtonian/non-baryonic contributions. The rotation curve of the BG general relativistic model of \cite{beordoGeometrydrivenDarkmattersustainedMilky2024e} is split between an effective Newtonian counterpart and a gravitational dragging contribution, which is a purely relativistic effect due to the spacetime geometry. The MWC and $\Lambda$CDM rotation curves are given by the quadratic summation of the baryonic and halo contributions, while the Mondian boost responsible for the flat profile is given by equation~(\ref{eq:v_mondBoost}).
From this plot, it is clear that the non-Newtonian/non-baryonic contributions onset between $10\text{--}15$ kpc, becoming dominant beyond 15 kpc. Due to the slightly higher baryonic mass in MOND, the baryonic component of the rotation curve tends to be slightly higher compared to the classical models with dark matter. Similarly, within the framework of $\Lambda$CDM, the dark matter halo described by the Einasto profile contributes more than the NFW case of the MWC model, where less dark matter and more baryonic matter are attributed.

\section{Conclusions}
In this study, we have undertaken a comprehensive analysis of Galactic rotation curves utilizing the latest Gaia Data Release 3, in line with seminal previous results with Gaia DR2 \cite{crostaTestingCDMGeometrydriven2020}. Considering larger samples from the latest Gaia Release, we compared the rotation curves within the frameworks of two prominent dynamical models: MOND and the $\Lambda$CDM paradigm. We extended the analysis previously presented in \cite{beordoGeometrydrivenDarkmattersustainedMilky2024e} with the same set of data, where we compared again the rotation curves using the BG and MWC models.

Our analysis revealed several key findings. Firstly, we found that both the MOND and $\Lambda$CDM models provided statistically equivalent fits to the Gaia DR3 rotation curve data across various stellar populations. This suggests that all four models are capable of accurately describing the observed dynamics of the Milky Way, albeit with different underlying assumptions.

The parameters of the baryonic matter distribution are consistent between the different datasets and models, although the $\Lambda$CDM paradigm tends to assign slightly less mass to the baryonic component compared to MOND and the MWC model. 

Furthermore, our analysis confirms the previously reported tension between our derived dynamical mass estimates and those reported in recent literature, claiming the presence of a Keplerian decline beyond 18-19 kpc. Our estimates of the virial mass range from $1.5\text{--}2.5 \times 10^{12} \, {\rm M}_{\odot}$ in the $\Lambda$CDM framework, which is an order of magnitude higher than proposed in the literature lately.
While our findings align with recent studies showing near-flat rotation curves within the observed range, we acknowledge the discrepancy and highlight the need for further investigation into the methodologies employed.

Additionally, we compared the contributions to the rotation curve from the baryonic and non-Newtonian components for each model. We found that the non-Newtonian/non-baryonic contributions become dominant beyond 10\text{--}15 kpc, with MOND predicting a slightly higher baryonic contribution compared to the classical models with dark matter, and $\Lambda$CDM attributing more dynamical mass to the dark matter halo described by the Einasto profile compared to the NFW halo in the MWC model.

Overall, our study underscores the importance of comparing different dynamical models in understanding the dynamics of galaxies. By leveraging the wealth of data provided by Gaia DR3, we have gained valuable insights into the structure and composition of our own Milky Way galaxy, shedding light on the behaviour and the validity of different theories of gravity. Future work should focus on resolving the discrepancies observed in the dynamical mass estimates and further exploring the implications for our understanding of the dark matter role and galactic structure, especially considering the subtle richness provided by a general relativistic scenario.

\section*{Data availability}
The data sets used in this analysis are presented by \cite{beordoGeometrydrivenDarkmattersustainedMilky2024e}. Posterior distributions or other data underlying this article will be shared on reasonable request to the corresponding author.

\appendix

\section{Jeans analysis of the sample}
\label{sec:jeans}
\begin{figure}[htbp]
    \centering
    \includegraphics[width=\linewidth]{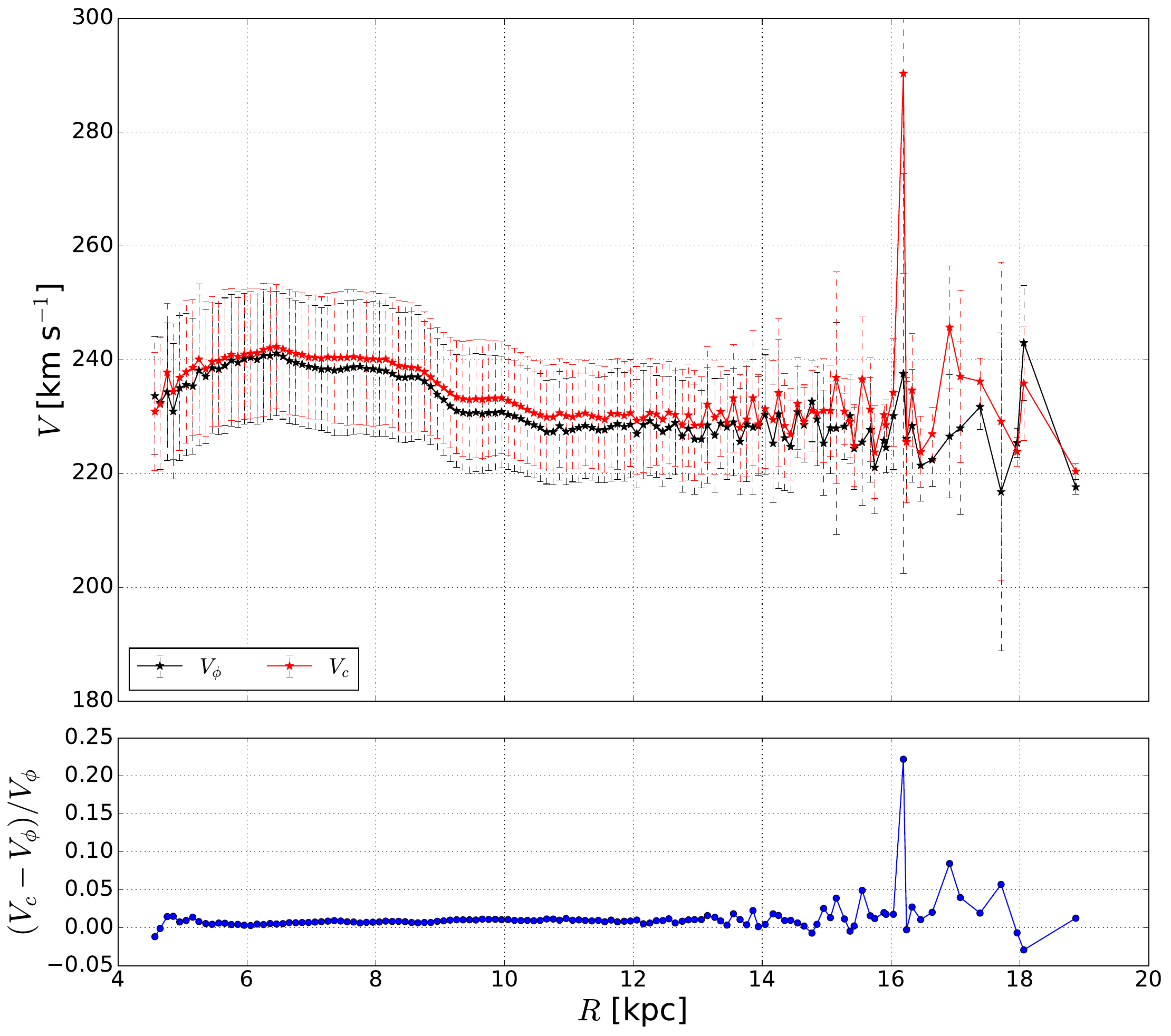}
    \caption{The top panel shows the circular velocity profile $V_c$ derived from the Jeans equation~(\ref{eq:jeans_final}) along with the azimuthal velocity $V_\phi$ for the full sample of stars. Here, $V_c$ and $V_\phi$ share the same error bars. The bottom panel highlights the relative difference between the two quantities.\label{fig:jeans}}
\end{figure}

Under the assumption of axisymmetric and stationary equilibrium, the Jeans equation in cylindrical coordinates is \cite{jeans1915, binneyGalacticDynamics2008}
\begin{equation}
    \dfrac{\partial(\nu \langle V_R^2 \rangle)}{\partial R} + \dfrac{\partial(\nu \langle V_R V_z \rangle)}{\partial z} + \nu \left( \dfrac{\langle V_R^2 \rangle - \langle V_\phi^2 \rangle}{R}  + \dfrac{\partial \phi}{\partial R} \right) = 0 \; ,
\end{equation}
where $\nu$ is the radial volume density of the Galaxy, $\phi$ is the gravitational potential, $V_R, V_\phi, V_z$ are the radial, azimuthal, and vertical velocities respectively, and the brackets $\langle \rangle$ represent quantities averaged over the velocity space.
Being the circular rotation velocity defined as
\begin{equation}
    V_c^2 = R \dfrac{\partial \phi}{\partial R}\biggr\rvert_{z \approx 0} \; ,
\end{equation}
the Jeans equation for $z \approx 0$ becomes 
\begin{equation}
    \Biggl[ \dfrac{\partial(\nu \langle V_R^2 \rangle)}{\partial R} + \dfrac{\partial(\nu \langle V_R V_z \rangle)}{\partial z} + \nu \left( \dfrac{\langle V_R^2 \rangle - \langle V_\phi^2 \rangle + V_c^2}{R} \right) \Biggr]_{z \approx 0} = 0 \; .
\end{equation}
Considering the vertical gradient of the cross term $\langle V_R V_z \rangle$ negligible, the equation can be written as
\begin{equation}
    V_c^2 = \left[ \langle V_\phi^2 \rangle - \langle V_R^2 \rangle \left( 1 + \dfrac{\partial \ln \nu}{\partial \ln R} + \dfrac{\partial \ln \langle V_R^2 \rangle}{\partial \ln R} \right) \right]_{z \approx 0} \; ,
    \label{eq:jeans_final}
\end{equation}
where the radial volume density is usually assumed to be $\nu \propto e^{-R/h_R}$. Widely used values for the radial scale length $h_R$ range from 2 to 5 kpc, therefore we adopt a value of 2.5 kpc \cite{juric2008, jiaoDetectionKeplerianDecline2023}. 
The radial gradient of the averaged squared radial velocity $\langle V_R^2 \rangle$, i.e. the last term in equation~(\ref{eq:jeans_final}), can instead be derived directly from the data set. Fitting $\sqrt{\langle V_R^2 \rangle}$ with an exponential function, we estimate a scale length of $\approx 26$ kpc, in line with other studies \cite{ouDarkMatterProfile2024, zhouCircularVelocityCurve2023, eilersCircularVelocityCurve2019}. The resulting circular velocities for the full sample are plotted in figure~\ref{fig:jeans}. These values typically exceed the azimuthal velocities by less than 5 per cent and fall well within the error bars, as expected, given that the orbital eccentricity selection performed in \cite{beordoGeometrydrivenDarkmattersustainedMilky2024e} has already cleared out most of the asymmetric drift.

\section{Dependence on the width of radial bins}
\label{sec:bins}
The choice of the radial bin size strongly affects the derived rotation curve, the more the size of the bin is large. This is because some information is lost and some features of the rotation curve are smoothed out by averaging the velocities of stars with different true radial coordinates, especially in less populated bins (i.e. at large radii where data points are crucial in the dynamical mass determination).
For the optimal choice of the bin size, in \cite{beordoGeometrydrivenDarkmattersustainedMilky2024e} we adopted Knuth's Rule \cite{knuthOptimalDataBasedBinning2006}, a data-based Bayesian algorithm, yielding to 0.1 kpc bins (0.5 kpc for the DCEP sample). However, in the literature, observed data are usually grouped in $0.5\text{--}1$ kpc radial bins.

\begin{figure}[htbp]
     \begin{subfigure}{.495\linewidth}
         \includegraphics[width=\linewidth]{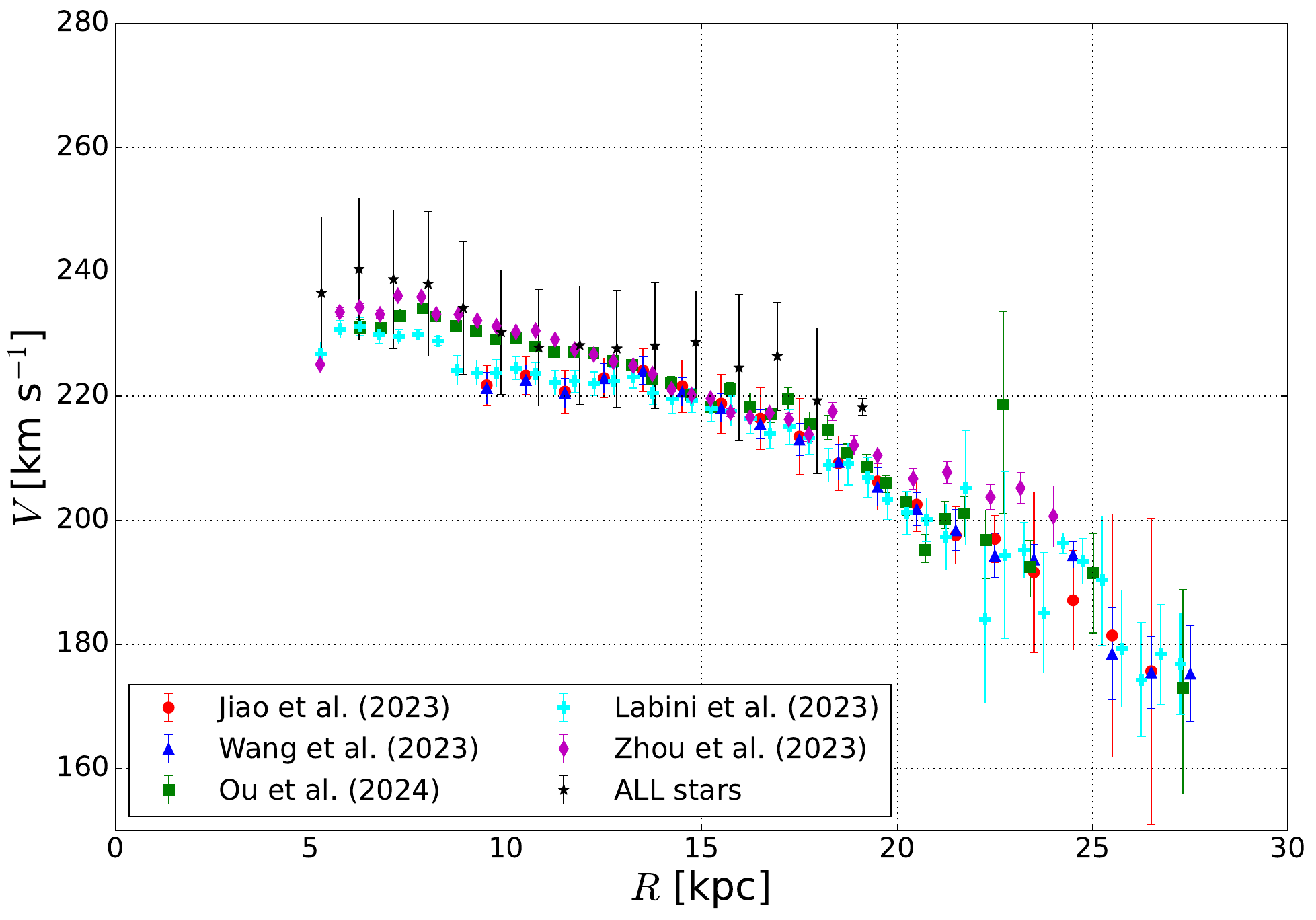}
     \end{subfigure}
     \hfill
     \begin{subfigure}{.495\linewidth}
         \includegraphics[width=\linewidth]{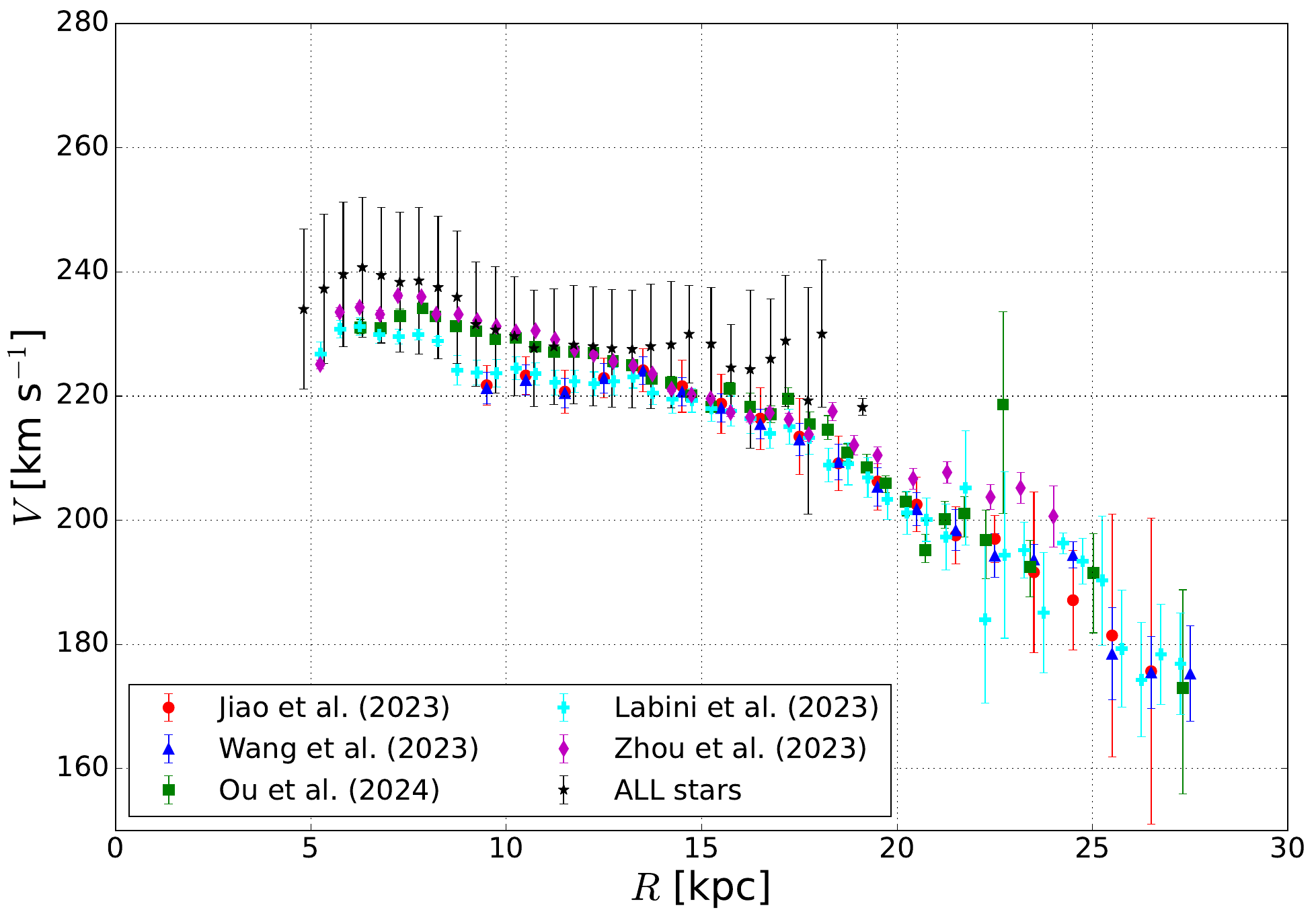}
     \end{subfigure}
    \caption{Same as figure~\ref{fig:comparison} but the azimuthal velocities of the full sample are averaged over radial bins of size equal to 1 and 0.5 kpc for the left and right panels respectively.\label{fig:different_bins}}
\end{figure}

Imposing radial bins of 1 kpc, we find the rotation curve shown in the top panel of figure~\ref{fig:different_bins}. The flattening around 15 kpc, although less clear, is still present; a similar but softer behaviour seems to be found by \cite{ouDarkMatterProfile2024} and \cite{zhouCircularVelocityCurve2023} at slightly larger distances and lower velocities. 
The bottom panel of the same figure shows the rotation curve obtained with 0.5 kpc radial bins instead, highlighting how data points are differently distributed with respect to the 1 kpc case.
However, if error bars are appropriately selected, one should obtain consistent results. In fact, when repeating the fitting process using the rotation curve derived with either 0.5 or 1 kpc bin width for the four models, we obtain estimates of the model parameters that are perfectly consistent with previous ones. This underlines the robustness of the rotation curve defined in \cite{beordoGeometrydrivenDarkmattersustainedMilky2024e} and used in the present work. On the contrary, it remains unclear to what extent the results are influenced by the radial bin size when error bars are derived via bootstrapping instead, necessitating further verification.

\acknowledgments
We wish to thank Paola Re Fiorentin and Alessandro Spagna for the selection of the stellar samples performed in \cite{beordoGeometrydrivenDarkmattersustainedMilky2024e}.
This work has made use of data products from: the ESA Gaia mission (gea.esac.esa.int/archive/), funded by national institutions participating in the Gaia Multilateral Agreement.%; and the Two Micron All Sky Survey (2MASS,www.ipac.caltech.edu/2mass).
We are indebted to the Italian Space Agency (ASI) for their continuing support through contract %2014-025-R.1.2015 to INAF. 
% supported by the Italian Space Agency (ASI) through contract 
2018-24-HH.0 and its addendum 2018-24-HH.1-2022 to the National Institute for Astrophysics (INAF).

\bibliographystyle{JHEP}
\bibliography{main}{}

\end{document}